\documentclass[conference]{IEEEtran}
\usepackage{amsfonts}
\IEEEoverridecommandlockouts

\ifCLASSINFOpdf
\else
\fi
\usepackage{epsfig}
\usepackage{graphicx}
\usepackage{psfig}
\usepackage{epsf}
\usepackage[cmex10]{amsmath}
\usepackage{booktabs}
\usepackage{fancyhdr}
\usepackage{color}
\usepackage{slashbox}
\usepackage{multirow}
\usepackage{caption2}   
\newtheorem{myTheo}{Theorem}   

\newtheorem{lemma}{\text{Lemma}}
\newtheorem{rem}{Remark} 

\begin{document}
\title{Computation Rate Maximization in UAV-Enabled Wireless Powered  Mobile-Edge Computing Systems}
\author{\IEEEauthorblockN{Fuhui Zhou, \emph{Member, IEEE}, Yongpeng Wu, \emph{Senior Member, IEEE},\\ Rose Qingyang Hu, \emph{Senior Member, IEEE}, and Yi Qian, \emph{Senior Member, IEEE}}

\thanks{
Manuscript received January 4, 2018; revised May 1, 2018 and
accepted June 4, 2018. Date of publication ****; date of
current version ****. The research of F. Zhou was supported in part by the Natural Science Foundation of China under Grant 61701214, in part by the Young Natural Science Foundation of Jiangxi Province under Grant 20171BAB212002, in part by The Open Foundation of The State Key Laboratory of Integrated Services Networks under Grant ISN19-08, and in part by The Postdoctoral Science Foundation of Jiangxi Province under Grant 2017M610400, Grant 2017KY04 and Grant 2017RC17. The research of Y. Wu was supported by the Natural Science Foundation of China under Grant 61701301 and in part by Young Elite Scientist Sponsorship Program by CAST. The research of Prof. R. Q. Hu was supported in part by the National Science Foundation under Grants EECS-1308006, NeTS-1423348, EARS-1547312 and the Natural Science Foundation of China under Grant 61728104. The research of Prof. Y. Qian was supported by the National Science Foundation under Grants EECS-1307580, NeTS-1423408 and EARS-1547330. The corresponding author is Yongpeng Wu.}
\thanks{F. Zhou is with the Department of Electrical and Computer Engineering as a Research Fellow at Utah State University, U.S.A. F. Zhou is also with the School of Information Engineering, Nanchang University, P. R. China, 330031. He is also with State Key Laboratory of Integrated Services Networks, Xidian University, Xi¡¯an,
710071, P. R. China (e-mail: zhoufuhui@ieee.org).

Y. Wu is with Shanghai Key Laboratory of Navigation and Location Based Services, Shanghai Jiao Tong University, Minhang, 200240, China (Email:yongpeng.wu2016@gmail.com).

R. Q. Hu is with the Department of Electrical and Computer Engineering, Utah State University, USA.  (e-mail: rose.hu@usu.edu).

Y. Qian is with the Department of Electrical and Computer Engineering, University of Nebraska-Lincoln, Omaha, NE 68182, USA. (E-mail: yqian2@unl.edu).
}

}
\maketitle
\begin{abstract}
Mobile edge computing (MEC) and wireless power transfer (WPT) are two promising techniques to enhance the computation capability and to prolong the operational time of low-power wireless devices that are ubiquitous in Internet of Things. However, the computation performance and the harvested energy are significantly  impacted by the severe propagation loss. In order to address this issue, an unmanned aerial vehicle (UAV)-enabled MEC wireless powered system is studied in this paper.  The computation rate maximization problems in a UAV-enabled MEC wireless powered system are investigated under both partial and binary computation offloading modes, subject to the energy harvesting causal constraint and the UAV's speed constraint. These problems are non-convex and challenging to solve. A two-stage algorithm and a three-stage alternative algorithm are respectively proposed for solving the formulated problems. The closed-form expressions for the optimal central processing unit frequencies,  user offloading time,  and user  transmit power  are derived. The optimal selection scheme on whether users choose to locally compute or offload computation tasks is proposed for the binary computation offloading mode. Simulation results show that our proposed resource allocation schemes outperforms other benchmark schemes. The results also demonstrate that the proposed schemes converge fast and  have low computational complexity.
\end{abstract}
\IEEEpeerreviewmaketitle
\section{Introduction}
\IEEEPARstart{T}{HE} Internet of Things (IoT) has been widely developed with the unprecedented proliferation of mobile devices, such as smart phones, cloud-based mobile sensors, tablet computers and wearable devices, which facilitates the realization of smart environment (e.g. smart city, smart home, smart transportation, etc.) \cite{F. Zhou4}. IoT enables mobile users to experience intelligent applications (e.g., automatic navigation, face recognition, unmanned driving, etc.) and to enjoy diverse services with high quality of service (QoS) such as  mobile online gaming, augmented reality, etc. These services normally require a massive number of size-constrained and low-power mobile devices to perform computation-intensive and latency-sensitive tasks \cite{Y. Mao}. However, it is challenging  for mobile devices to perform these services due to their low computing capability and finite battery lifetime.

Mobile edge computing (MEC) and wireless power transfer (WPT) have been deemed two promising technologies to tackle the above mentioned challenges \cite{Y. Mao}-\cite{F. Zhou2}. Recently, MEC has received an ever-increasing level of attention from industry and academia since it can significantly improve the computation capability of mobile devices in a cost-effective and energy-saving manner \cite{Y. Mao}. It enables mobile devices to offload partial or all of their computation-intensive tasks to MEC servers that locate at the edge of the wireless network, such as cellular base stations (BSs) and access points (APs). Different from the conventional cloud computing, MEC  servers are deployed in a close proximity to end users. Thus, MEC has the potential to provide low-latency services, to save energy for mobile users, and to achieve high security \cite{Y. Mao}. Up to now, there are a number of leading companies (e.g.,  IBM, Intel, and Huawei) that have identified MEC as a promising technique for  the future wireless communication networks. In general, MEC has two operation modes, namely, partial and binary computation offloading.  In the first mode, the computation task can be partitioned into two parts, and one part is locally executed  while the other part is offloaded to the MEC servers for computing \cite{S. Sardellitti}-\cite{L. Liu}. For the second mode,  computation tasks cannot be partitioned. Thus they can be either executed locally or completely offloaded  \cite{W. Zhang}.

On the other hand, WPT can provide low-power mobile devices with sustainable and cost-effective energy supply by using radio-frequency (RF) signals \cite{X. Lu}. It facilitates a perpetual operation and enables users to have high QoE, especially in the case that mobile devices do not have sufficient battery energy for offloading task or taking the services when the battery energy is exhausted. Compared to the conventional energy harvesting techniques, such as solar or wind charging, WPT is more attractive since it can provide a controllable and stable power supply \cite{F. Zhou2}. It is envisioned that the computation performance can be significantly improved by integrating WPT into MEC networks \cite{J. Xu}-\cite{S. Bi}. However, the harvested power level can be significantly degraded by the severe propagation loss. Recently, an unmanned aerial vehicle (UAV)-enabled WPT architecture has been proposed to improve the energy transfer efficiency \cite{H. Wang}-\cite{J. Xu3}. It utilizes an unmanned aerial vehicle (UAV) as an energy transmitter for powering the ground mobile users. It was shown that the harvested power level can be greatly  improved due to the fact that there is a  high possibility  that short-distance line-of-sight (LoS) energy transmit links exist \cite{H. Wang}-\cite{J. Xu3}. Moreover, the computation performance can also be improved by using the UAV-assisted MEC  architecture \cite{N. H. Motlagh}-\cite{M. A. Messous}. Furthermore, UAV-assisted architectures can provide flexible deployment and low operational costs, and are particularly  helpful in the situations that the conventional communication systems are destroyed by  natural disasters \cite{Y. Zeng1}-\cite{C. X. Wang}.

Motivated by the above mentioned reasons, a UAV-enabled and wireless powered MEC network is studied in this paper. In order to maximize the achievable computation rate, the communication and computation resources and the trajectory of the UAV are jointly optimized under both partial and binary computation offloading modes. To the authors' best knowledge, this is the first work that considers the UAV-enabled  wireless powered MEC network and studies the computation rate maximization problems in this type of network.
\subsection{Related Work and Motivation}
In wireless powered MEC systems, it is of great importance to design resource allocation schemes so as to efficiently exploit energy, communication, and computation resources and improve the computation performance. Resource allocation problems have been extensively investigated in the conventional MEC networks \cite{S. Sardellitti}-\cite{W. Zhang} and also in MEC networks relying on energy harvesting \cite{J. Xu}-\cite{S. Bi}. Recently, efforts have also been dedicated to designing resource allocation and trajectory schemes in UAV-enabled wireless powered communications network \cite{ H. Wang}-\cite{J. Xu3} and UAV-assisted MEC networks \cite{N. H. Motlagh}-\cite{M. A. Messous}. These contributions are summarized as follows.

In MEC networks, the communication and computation resources and the selection of the offloading mode were jointly optimized to achieve the objective of the system design, e.g., the users' consumption energy minimization \cite{ S. Sardellitti}, \cite{C. You1}, the revenue maximization \cite{C. Wang}, the maximum cost minimization \cite{J. Du}, etc. Specifically,  in \cite{ S. Sardellitti}, the total energy of all users in a multi-cell MEC network was minimized by jointly optimizing the user transmit precoding matrices  and the central processing unit (CPU) frequencies of the MEC server allocated to each user. It was shown that the performance achieved by jointly optimizing the communication and computation resources is superior to that obtained by  optimizing these resources separately. The authors in \cite{C. You1} extended the energy minimization problem into the multi-user MEC systems with time-division multiple access (TDMA) and orthogonal frequency-division multiple access (OFDMA), respectively. It was proved that the optimal offloading policy has a threshold-based structure, which is related to the channel state information (CSI) \cite{C. You1}. Particularly, mobile users offload their computation tasks when the channel condition is strong; otherwise, they can locally execute the computation tasks. In \cite{C. Wang}, the revenue of the wireless cellular networks with MEC was maximized by jointly designing the computation offloading decision, resource allocation, and content caching strategy. The works in \cite{ S. Sardellitti}-\cite{C. Wang} focused on optimizing a single objective, which over-emphasizes the importance of one metric and may not  achieve a good tradeoff among multiple metrics. Recently, the authors in \cite{J. Du} and \cite{L. Liu} studied  the fairness and multi-objective optimization problem in MEC networks. It was shown that there exist multiple tradeoffs in MEC systems, such as the tradeoff between the total computation rate and the fairness among users. Different from the works in \cite{ S. Sardellitti}-\cite{L. Liu}, MEC systems with the binary computation offloading mode were considered and the optimal resource allocation strategy was designed to minimize the consumption energy in \cite{W. Zhang}.

Energy harvesting was not considered in the MEC systems \cite{S. Sardellitti}-\cite{W. Zhang}. Recently, the authors in \cite{J. Xu}-\cite{S. Bi} have studied the resource allocation problem in various MEC systems relying on energy harvesting. In \cite{J. Xu} and \cite{Y. Mao1}, The reinforcement learning and Lyapunov optimization theory were used to design resource allocation schemes in MEC systems relying on the conventional energy harvesting techniques. Different from \cite{J. Xu} and \cite{Y. Mao1}, the resource allocation problems were studied in wireless powered MEC systems \cite{C. You}-\cite{S. Bi}. Specifically, the authors in \cite{C. You} proposed an energy-efficient computing framework in which the energy consumed for local computing and task offloading is from the harvested energy. The consumed energy was minimized by jointly optimizing the CPU frequency and the mode selection.  In \cite{F. Wang}, the  energy minimization problem was extended into a multi-input single-out wireless powered MEC system, and the offloading time, the offloading bits, the CPU frequency and the energy beamforming were jointly optimized. Unlike \cite{F. Wang}, energy efficiency was defined and maximized in a full-duplex wireless powered MEC system by jointly optimizing the transmission power, offloaded bits, computation energy consumption, time slots for computation offloading and energy transfer \cite{S. Mao}. In contrast to the work in \cite{C. You}-\cite{S. Mao}, the computation bits were maximized in a wireless powered MEC system under the binary computation offloading mode \cite{S. Bi}.  Two sub-optimal algorithms based on the alternating direction method were proposed to solve the combinatorial programming problem.  The  proposed algorithms actually did not provide the optimal selection scheme for the user operation mode.

Although WPT has been exploited to improve the computation performance of MEC systems \cite{C. You}-\cite{S. Bi}, the energy harvested by using WPT can be significantly degraded  by the severe propagation loss.  The energy conversion efficiency is  low when the distance between the energy transmitter and the harvesting users is large.  In order to tackle this challenge, the authors in \cite{H. Wang}-\cite{ J. Xu3} proposed a UAV-enabled wireless powered architecture where a UAV transmits energy to the harvesting users. Due to the high possibility of having line-of-sight (LoS) air-to-ground energy harvesting links, the harvesting energy can be significantly improved by using this architecture. Moreover, it was shown that the harvesting energy can be further improved by optimizing the trajectory of the UAV \cite{ S. Yin}-\cite{ J. Xu3}. Thus, it is envisioned that the application of the UAV-enabled architecture into wireless powered MEC systems  is promising and valuable to be studied \cite{Y. Zeng1}. However, to the authors' best knowledge, few investigations have focused on this area.

Recently, the UAV-enabled MEC systems have been studied and their resource allocation schemes have been proposed \cite{ N. H. Motlagh}-\cite{M. A. Messous}. In \cite{ N. H. Motlagh}, the UAV-enabled MEC architecture was first proposed and  the computation performance was improved by using UAV. The authors in \cite{N. Zhao} proposed a new caching UAV framework to help small cells to offload traffic. It was shown that the throughput can be greatly improved  while the overload of wireless backhaul can be significantly reduced. In order to further improve the computation performance, the authors in \cite{S. Jeong} and \cite{ S. Jeong2} designed a resource allocation scheme that jointly optimizes the CPU frequency and the trajectory of the UAV. In \cite{M. A. Messous}, a theoretical game method was applied to design a resource allocation scheme for the UAV-enabled MEC system and the existence of Nash Equilibrium was demonstrated.

Although resource allocation problems have been well studied in MEC systems \cite{S. Sardellitti}-\cite{W. Zhang},  MEC systems relying on energy harvesting \cite{J. Xu}-\cite{S. Bi} and UAV-enabled MEC systems \cite{N. H. Motlagh}-\cite{M. A. Messous}, few investigations have been conducted for designing resource allocation schemes in the UAV-enabled wireless powered MEC systems. Moreover, resource allocation schemes proposed in the above-mentioned works are inappropriate to UAV-enabled MEC wireless powered systems since the computation performance not only depends on the optimization of energy, communication and computation resources, but also relies on the design of the UAV trajectory. Furthermore, the application of UAV into wireless powered MEC systems  has the potential to enhance the user computation capability since it can improve the energy conversion efficiency and task offloading efficiency \cite{N. Cheng}, \cite{M. Mozaffari}. Thus, in order to improve the computation performance and provide mobile users with high QoE, it is of great importance and  worthiness to study resource allocation problems in  UAV-enabled wireless powered MEC systems. However, these problems are indeed  challenging to tackle. The reasons are from two aspects. On one hand, there exists dependence among different variables (e.g., the CPU frequency, the task offloading time and the variables related to the trajectory of the UAV), which makes the problems non-convex. On the other hand, when the binary computation offloading mode is applied, the resource allocation problems in UAV-enabled wireless powered MEC systems have binary variables related to the selection of either local computation or offloading tasks. It makes the problem a  mixed integer non-convex optimization problem.
\subsection{Contributions and Organization}
In contrast to \cite{S. Sardellitti}-\cite{S. Bi}, this paper studies the resource allocation problem in UAV-enabled wireless powered MEC systems, where a UAV transmits energy signals to charge multiple mobile users and provides computation services for them.  Although the computation performance is limited by the flight time of the UAV, it is worth studying UAV-enabled wireless powered MEC systems since these systems  are promising in environments such as mountains and desert areas, where no terrestrial wireless infrastructures  exist,  and in environments where the terrestrial wireless infrastructures are destroyed due to the natural disasters \cite{N. Cheng}, \cite{M. Mozaffari}. Thus, in this paper, the weighted sum  computation bits of all users are maximized under both partial and binary computation offloading modes. The main contributions of this work are summarized as follows:
\begin{enumerate}
  \item It is the first time that the resource allocation framework is formulated in UAV-enabled MEC wireless powered systems under both  partial and binary computation offloading modes. The weighted sum computation bits are  maximized by jointly optimizing the CPU frequencies, the offloading times and the transmit powers of users as well as  the UAV trajectory. Under the partial computation offloading mode, a two-stage alternative algorithm is proposed to solve the non-convex and challenging computation bits maximization problem. The closed-form expressions for the optimal CPU frequencies, the offloading times and the transmit powers of users are derived for any given trajectories.
  \item Under the binary computation offloading mode, the weighted sum computation bits maximization problem is a mixed integer non-convex optimization problem, for which a three-stage alternative algorithm is proposed. The optimal selection scheme on  whether users choose to locally compute or offload tasks is derived in a closed-form expression for a given trajectory. The structure for the optimal selection scheme shows that whether users choose to locally compute or offload their tasks to the UAV for computing depends on the tradeoff between the achievable computation rate and the operation cost. Moreover, the trajectory of the UAV is optimized by using the successive convex approximation (SCA) method under both  partial and binary computation offloading modes.
  \item The simulation results show that the computation performance obtained by using the  proposed resource allocation scheme is better than  these achieved by using the disjoint optimization schemes. Moreover, it only takes several iterations for  the proposed alternative algorithms to converge. Furthermore, simulation results verify that the priority and fairness of users can be improved by using the weight vector. Additionally, it is shown that the total computation bits increase with the number of users.
\end{enumerate}

The remainder of this paper is organized as follows. Section II gives the system model. The resource allocation problem  is formulated under the partial computation offloading mode in Section III. Section IV formulates the resource allocation problem under the binary computation offloading mode. Simulation results are presented in Section V. Finally, our paper is concluded in Section VI.
\section{System Model}
\begin{figure}[!t]
\centering
\includegraphics[width=3.5 in]{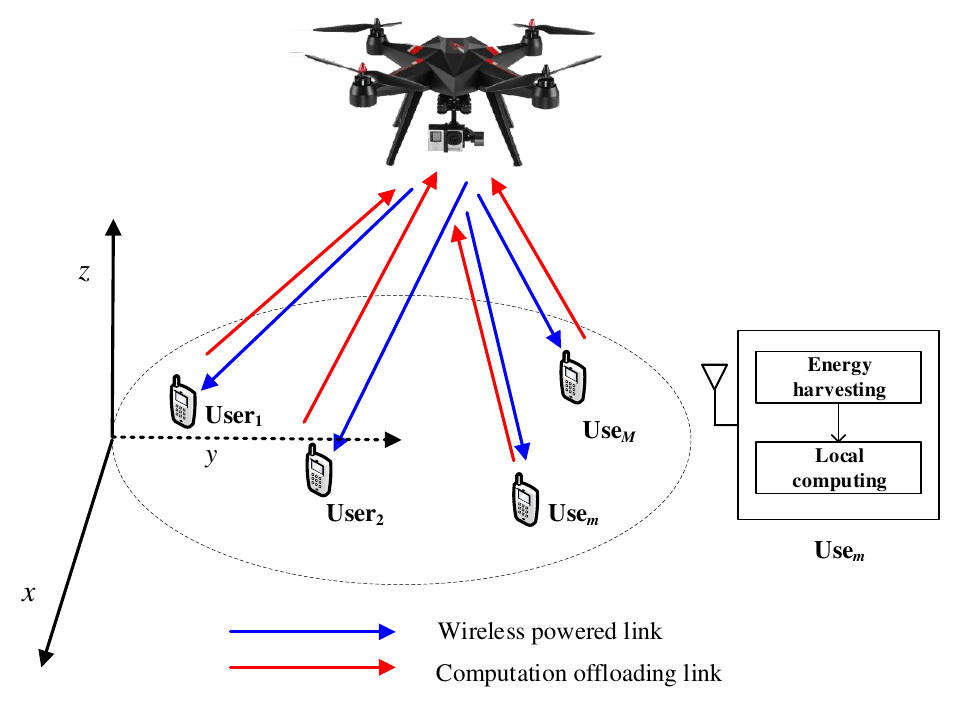}
\caption{The system model.} \label{fig.1}
\end{figure}
A UAV-enabled wireless powered MEC system is considered in Fig. 1, where an RF energy transmitter and an MEC
server are implemented in UAV. The UAV transmits energy to $M$ users and provides MEC services for these users. Each user has an energy harvesting circuit and can store energy for its operation. The UAV has an on-board communication circuit and an on-board computing processor. So does each user.  The computing processor of each user is an on-chip micro-processor that has low computing capability and can locally execute simple tasks. The UAV has a powerful processor that can perform computation-intensive tasks \cite{N. H. Motlagh}-\cite{ M. A. Messous}. Similar to \cite{C. You}-\cite{S. Bi}, each user can simultaneously perform energy harvesting, local computing and computation offloading while the UAV can simultaneously  transmit energy and perform computation. In this paper, all devices are equipped with a single antenna.

Without loss of generality, a three-dimensional (3D) Euclidean coordinate is adopted. Each user\rq{}s location  is fixed on the ground. The location of the $m$th ground user is denoted by $\mathbf{q}_m$, where $\mathbf{q}_m=[x_m, y_m]$, $m\in {\cal M}$ and $ {\cal M}=\left\{1, 2, \cdots, M\right\}$. Boldface lower case letters represent vectors and boldface upper case letters represent matrices. $x_m$ and $y_m$ are the horizontal plane coordinates of the $m$th ground user. It is assumed that user positions are known to the UAV for designing the trajectory \cite{ S. Yin}-\cite{ J. Xu3}. A finite time horizon with duration $T$ is considered. During $T$, the UAV flies at the same altitude level denoted by $H$ ($H>0$). In practice, the fixed altitude is the minimum altitude that is appropriate to the work terrain and can avoid building without the requirement of frequent aircraft descending and ascending. A block fading channel model is applied, i.e., during each $T$,  the channel  remains static.

For the ease of exposition, the finite time $T$ is discretized into $N$ equal time slots, denoted by $n= 1, 2, \cdots, N$. At the $n$th slot, it is assumed that the horizontal plane coordinate of the UAV is $\mathbf{q}_u\left[n\right]=[x_u[n], y_u[n]]$. Similar to \cite{Y. Zeng}-\cite{C. X. Wang}, it is assumed that the wireless channel between the UAV and each user is dominated by LOS. Thus, the channel power gain between the UAV and the $m$th user, denoted by ${h_m}\left[ n \right]$, can be given as
\begin{align}\label{27}\
{h_m}\left[ n \right] = {\beta _0}d_{m,n}^{ - 2} = \frac{{{\beta _0}}}{{{H^2} + {{\left\| {{\mathbf{q}_u}\left[ n \right] - {\mathbf{q}_m}} \right\|}^2}}}, m\in {\cal M}, n\in {\cal N},
\end{align}
where $\beta _0$ is the channel power gain at a reference distance $d_0=1$ m; $d_{m,n}$ is the horizontal plane distance between the UAV and the $m$th user at the $n$th slot, $n\in {\cal N}$, ${\cal N}=\left\{1, 2, \cdots, N\right\}$;  $\left\| \cdot\right\|$ denotes its Euclidean norm. The details for the UAV-enabled wireless powered MEC system are presented under partial and binary computation offloading modes in the following, respectively.
\subsection{Partial Computation Offloading Mode}
Under the partial computation offloading mode, the computation task of each user can be partitioned into two parts, one for local computing and one for offloading to the UAV. The energy consumed for local computing and task offloading comes from the harvested energy. In this paper, in order to shed meaningful insights into the design of a UAV-enabled wireless powered MEC system, similar to \cite{F. Zhou2}, \cite{C. You}-\cite{S. Bi}, the linear energy harvesting model is applied. Thus, the harvested energy ${E_m}\left[ n \right]$ at the $m$th user during  $n$  time slots is given as
\begin{align}\label{27}\
{E_m}\left[ n \right] = \sum\limits_{i = 1}^n {\frac{{T\eta_0 {h_m}\left[ i \right]{P_0}}}{N}}, m\in {\cal M}, n\in {\cal N},
\end{align}
where $\eta_0$ denotes the energy conservation efficiency, $0<\eta_0\leq1$ and ${P_0}$ is the transmit power of the UAV. In this paper, the UAV employs a constant power transmission \cite{ S. Yin}-\cite{ J. Xu3}. The details for the operation of each user under the partial computation offloading mode are presented as follows.

\subsubsection{Local Computation} Similar to \cite{F. Wang}-\cite{ S. Bi}, the energy harvesting circuit, the communication circuit, and the computation unit are all separate. Thus, each user can simultaneously perform  energy harvesting, local computing,  and computation offloading. Let $C$ denote the number of CPU cycles required for computing one bit of raw data at each user. In order to efficiently use the harvested energy, each user adopts a dynamic voltage and frequency scaling technique and then can adaptively control the energy consumed for performing local computation by adjusting the CPU frequency during each time  slot \cite{F. Wang}-\cite{ S. Bi}. The CPU frequency of the $m$th user during the $n$th slot is denoted by ${{f_m}\left[ n \right]}$ with a unit of cycles per second.  Thus, the total computation bits executed at the $m$th user during $n$ slots and the total consumed energy at the $m$th user during $n$ slots are respectively given as ${\sum\limits_{k = 1}^n {\frac{{T{f_m}\left[ k\right]}}{{NC}}} }$ and $\sum\limits_{k = 1}^n { {{\gamma _c}f_m^3\left[ k \right] } }$ \cite{F. Wang}-\cite{ S. Bi}, where $\gamma _c$ is the effective capacitance coefficient of the processor's chip at the $m$th user,  $n\in {\cal N}$, $m\in {\cal M}$. Note that $\gamma _c$  is dependent of the chip architecture of the $m$th user.

\begin{figure}[!t]
\centering
\includegraphics[width=3.5 in]{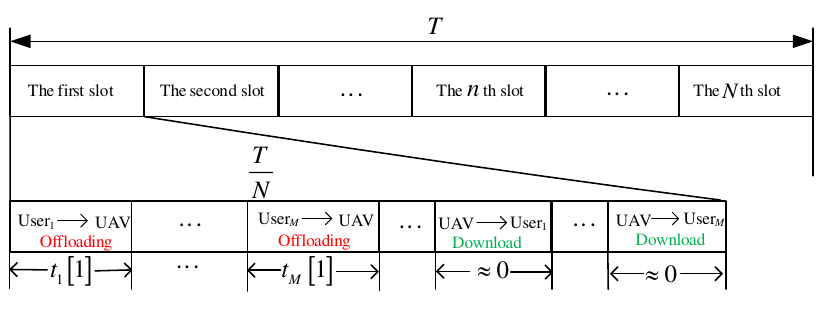}
\caption{The TDMA protocol for multiuser computation offloading.} \label{fig.1}
\end{figure}
\subsubsection{Computation Offloading} In order to avoid interference among users during the offloading process, a TDMA protocol shown in Fig. 2 is applied. Specifically, each time slot consists of three stages, namely, the offloading stage, the computation stage, and the downloading stage. In the offloading stage, $M$ users offload their respective computation task one by one during each slot. Let ${{t_m}\left[ n \right]} \times T/N$ $\left(0\leq {t_m}\left[ n \right]\leq1 \right)$ denote the duration in which  the $m$th user offloads its computation task to the UAV at the $n$th slot, $n\in {\cal N}$, $m\in {\cal M}$. Similar to \cite{ S. Bi}, the computation task of the $m$th user to be offload is composed of raw data and communication overhead, such as the encryption and packer header. Let ${{\nu _m}}{R_m}\left[ n \right]$ denote the total number of bits that the $m$th user offloads to the UAV during the $n$th slot, where ${R_m}\left[ n \right]$ is the number of raw data to be computed at the UAV  and $\nu _m$ indicates the communication overhead included in the offloading task. Thus, one has
\begin{align}\label{27}\ \notag
{R_m}\left[ n \right] \le \frac{{BT{t_m}\left[ n \right]}}{{{\nu _m}N}}{\log _2}\left( {1 + \frac{{{h_m}\left[ n \right]{P_m}\left[ n \right]}}{{\sigma _0^2}}} \right),\\ n\in {\cal N},m\in {\cal M},
\end{align}
where $B$ is the communication bandwidth; ${P_m}\left[ n \right]$ is the transmit power of the  $m$th user at the $n$th slot and $\sigma _0^2$ denotes the noise power at the $m$th user.

After all users offload their computation tasks at the $n$th slot, the UAV performs computing task and sends the computing results back to all the users. Similar to \cite{F. Wang}-\cite{ S. Bi}, the computation time and the downloading time of the UAV are neglected since the UAV has a much stronger computation capability than the users and the number of the bits related to the computation result is very small. Since the total offloading time of all users does not exceed the duration of one time slot, one has
\begin{align}\label{27}\
\sum\limits_{m = 1}^M {{t_m}\left[ n \right]}  \le 1, n\in {\cal N}.
\end{align}

Since the energy consumed for local computing and task offloading comes from the harvested energy, the following energy harvesting causal constraint should be satisfied.
\begin{align}\label{27}\ \notag
\frac{T}{N}\sum\limits_{k = 1}^n {\left[ {{\gamma _c}f_m^3\left[ k \right] + {t_m}\left[ k \right]{P_m}\left[ k \right]} \right]}  \le \frac{{{\eta _0}T}}{N}\sum\limits_{k = 1}^n {{h_m}\left[ k \right]{P_0}},\\ n\in {\cal N},m\in {\cal M}.
\end{align}

Under the partial computation offloading mode, the total computation bits $R_m$ of the $m$th user is given as
\begin{align}\label{27}\ \notag
&{R_m} = \sum\limits_{n = 1}^N {\frac{{T{f_m}\left[ n \right]}}{{NC}}}  + \frac{{BT{t_m}\left[ n \right]}}{{{\nu _m}N}}{\log _2}\left( {1 + \frac{{{h_m}\left[ n \right]{P_m}\left[ n \right]}}{{\sigma _0^2}}} \right),
\\ &\ \ \ \ \ \ \ \ \ \ \ \ \ \ \ \ \ \ \ \ \ \ \ \ \ \ \ \ \ \ \ \ \ \  m\in {\cal M}.
\end{align}
\subsection{Binary Computation Offloading Mode}
Under the binary computation offloading mode, the computation task cannot be partitioned. All the users need to choose to either  locally compute the task completely or offload the entire  task. This case can be widely experienced  in practice. For example, in order to improve the estimation accuracy, the raw data samples that are correlated need to be jointly computed altogether \cite{ W. Zhang}, \cite{ S. Bi}. Let ${\cal M}_0$ and ${\cal M}_1$ denote the set of  users that choose to perform local computation and the set of users that choose to perform task offloading, respectively. Thus, ${\cal M}={\cal M}_0\cup{\cal M}_1$ and ${\cal M}_0 \cap{\cal M}_1=\Theta$, where $\Theta$ denotes the null set.
\subsubsection{Users Choosing to Perform Local Computing} In this case, a user  in ${\cal M}_0$ exploits all the harvested energy to perform local computing. Thus, the total computation rate of the $i$th user denoted by $R_i^L$ can be given as
\begin{align}\label{27}\
R_i^L = \sum\limits_{n = 1}^N {\frac{{T{f_i}\left[ n \right]}}{{NC}}}, i\in {{\cal M}_0}.
\end{align}
And the energy harvesting causal constraint for a user in  ${\cal M}_0$ can be given as
\begin{align}\label{27}\
\frac{T}{N}\sum\limits_{k = 1}^n {{\gamma _c}f_i^3\left[ k \right]}  \le \frac{{{\eta _0}T}}{N}\sum\limits_{k = 1}^n {{h_i}\left[ k \right]{P_0}} , n\in {\cal N}, i\in {{\cal M}_i}.
\end{align}

\subsubsection{Users Choosing to Perform Task Offloading}  Each  user in ${\cal M}_1$ exploits all the harvested energy to perform task offloading. The TDMA protocol is applied to avoid interference among these users during the offloading process. Since the total offloading time of all users in ${\cal M}_1$  at the $n$th slot cannot exceed the duration of a time slot, one has
\begin{align}\label{27}\
\sum\limits_{j \in {{\cal M}_1}} {{t_j}\left[ n \right]}  \le 1, n\in {\cal N}.
\end{align}

Let $R_j^O$ denote the total computation rate of the $j$th user in the set ${\cal M}_1$. Then, one has
\begin{align}\label{27}\
R_j^O = \sum\limits_{n = 1}^N {\frac{{BT{t_j}\left[ n \right]}}{{{\nu _j}N}}{{\log }_2}\left( {1 + \frac{{{h_j}\left[ n \right]{P_j}\left[ n \right]}}{{\sigma _0^2}}} \right)}, j\in {{\cal M}_1}.
\end{align}
The energy harvesting causal constraint for a user in ${\cal M}_1$ can be given as
\begin{align}\label{27}\
\frac{T}{N}\sum\limits_{k = 1}^n {{t_j}\left[ k \right]{P_j}\left[ k \right]}  \le \frac{{{\eta _0}T}}{N}\sum\limits_{k = 1}^n {{h_j}\left[ k \right]{P_0}} , n\in {\cal N}, j\in {{\cal M}_1}.
\end{align}
Sections III and IV will respectively formulate the computation rate maximization problem for the  partial and binary computation offloading modes.
\section{Resource Allocation Under The Partial Computation Offloading Mode}
In this section, the resource allocation problem is studied under the partial computation offloading mode. The weighted sum computation bits are maximized by jointly optimizing the CPU frequencies, the offloading times and the transmit powers of users as well as the trajectory of the UAV. In order to tackle this non-convex  problem, a two-stage alternative algorithm is proposed.
\subsection{Resource Allocation Problem Formulation}
Under the partial computation offloading mode, the weighted sum computation bits maximization problem in the UAV-enabled wireless powered MEC system is formulated as $\mathbf{P}_{1}$,
\begin{subequations}
\begin{align}\label{27}\ \notag
&  \mathbf{P}_{1}: {\mathop {\max }\limits_{{f_m}\left[ n \right],{P_m}\left[ n \right],\mathbf{q}_u\left[ n \right],{t_m}\left[ n \right]} }\ {\sum\limits_{m = 1}^M {{w_m}\times} }\\
  &\left[ {\sum\limits_{n = 1}^N {\frac{{T{f_m}\left[ n \right]}}{{NC}}}  + \frac{{BT{t_m}\left[ n \right]}}{{{\nu _m}N}}{{\log }_2}\left( {1 + \frac{{{h_m}\left[ n \right]{P_m}\left[ n \right]}}{{\sigma _0^2}}} \right)} \right] \\
 &\text{s.t.}\ C1:{f_m}\left[ n \right]\geq0, {P_m}\left[ n \right]\geq0, m\in {\cal M},n\in {\cal N},\\ \notag
&C2:\frac{T}{N}\sum\limits_{k = 1}^n {\left[ {{\gamma _c}f_m^3\left[ k \right] + {t_m}\left[ k \right]{P_m}\left[ k \right]} \right]}  \le \frac{{{\eta _0}T}}{N}\sum\limits_{k = 1}^n {{h_m}\left[ k \right]{P_0}}\\
& \ \ \ \ \ \ \ \ \ \ \ \ \ \ \ \ \ \ \ \ \ \ \ \ \ \ \ \ \ \ \ \ \ \ \ \ \  \ m\in {\cal M},n\in {\cal N},\\
&C3:\sum\limits_{m = 1}^M {{t_m}\left[ n \right]}  \le 1,n\in {\cal N},\\
&C4:{\left\| {\mathbf{q}_u\left[ {n + 1} \right] - \mathbf{q}_u\left[ n \right]} \right\|^2} \le {V_{\max }}\frac{T}{N},n\in {\cal N},\\
&C5:\mathbf{q}_u\left[ 1 \right] = {\mathbf{q}_0},\mathbf{q}_u\left[ {N + 1} \right] = {\mathbf{q}_F},
\end{align}
\end{subequations}
where $V_{\max }$ denotes the maximum speed of the UAV in the unit of meter per second; $\mathbf{q}_0$ and $\mathbf{q}_F$ are the  initial and final  horizontal locations of the UAV, respectively. In $\left(12\right)$, $w_m$ denotes the weight of the $m$th user, which takes the priority and the fairness among users into consideration. $C1$ is the CPU frequency constraint and the computation offloading power constraint imposed on each user;  $C2$ represents the energy harvesting causal constraint;  $C3$ is the time constraint that the total time of all users offloading the computation bits cannot exceed the duration of each time slot;  $C4$ and $C5$ are the speed constraint and the initial and final  horizontal location constraint of the UAV, respectively. $\mathbf{P}_{1}$ is non-convex since there exist non-linear couplings among the variables, ${f_m}\left[ n \right]$, ${P_m}\left[ n \right]$,$\mathbf{q}_u\left[ n \right]$, ${t_m}\left[ n \right]$ and the objective function is non-concave with respect to the trajectory of the UAV.  In order to solve it, a two-stage alternative optimization algorithm is proposed. The details for the algorithm are presented as follows.
\subsection{Two-Stage Alternative Optimization Algorithm}
Let ${z_m}\left[ n \right] = {t_m}\left[ n \right]{P_m}\left[ n \right], n\in {\cal N}$. For a given trajectory, $\mathbf{P}_{1}$ can be transformed into $\mathbf{P}_{2}$.
\begin{subequations}
\begin{align}\label{27}\ \notag
&  \mathbf{P}_{2}: {\mathop {\max }\limits_{{f_m}\left[ n \right],{z_m}\left[ n \right],{t_m}\left[ n \right]} }\ {\sum\limits_{m = 1}^M {{w_m}} }\\
  & \times\left[ {\sum\limits_{n = 1}^N {\frac{{T{f_m}\left[ n \right]}}{{NC}}}  + \frac{{BT{t_m}\left[ n \right]}}{{{\nu _m}N}}{{\log }_2}\left( {1 + \frac{{{h_m}\left[ n \right]{z_m}\left[ n \right]}}{{{t_m}\left[ n \right]\sigma _0^2}}} \right)} \right]\\
\text{s.t.}\ & C1, C3,\\ \notag
&C5:\frac{T}{N}\sum\limits_{k = 1}^n {\left[ {{\gamma _c}f_m^3\left[ k \right] + {z_m}\left[ k \right]} \right]}  \le \frac{{{\eta _0}T}}{N}\sum\limits_{k = 1}^n {{h_m}\left[ k \right]{P_0}} ,\\
& \ \ \ \ \ \ \ \ \ \ \ \ \ \ \ \ \ \ \ \ \ \ \ \ \ \ \ \ \ \ \ \ \ \ \ \ \ \ m\in {\cal M},n\in {\cal N}.
\end{align}
\end{subequations}
It is easy to prove that $\mathbf{P}_{2}$ is convex and can be solved by using the Lagrange
duality method \cite{S. P. Boyd}, based on which the optimal solutions for the CPU frequency and the transmit power can be derived. Let $f_m^{opt}\left[ n \right]$ and $P_m^{opt}\left[ n \right]$ denote the optimal CPU frequency and transmit power of the $m$th user at the $n$th time slot, respectively, where $m\in {\cal M}$ and $n\in {\cal N}$. By solving $\mathbf{P}_{2}$, Theorem 1 can be stated as follows.
\begin{myTheo}
For a given trajectory $\mathbf{q}_u\left[ n \right]$, the optimal CPU frequency and transmit power of users can be respectively expressed as
\begin{subequations}
\begin{align}\label{27}\
&f_m^{opt}\left[ n \right] = \sqrt {\frac{{{w_m}}}{{3C{\gamma _c}\sum\limits_{k = n}^N {{\lambda _{m,k}}} }}},\\
&P_m^{opt}\left[ n \right]{\rm{ = }}\left\{ \begin{array}{l}
\begin{array}{*{20}{c}}
{0,}&{\text{if}\ {t_m}\left[ n \right] = 0,}
\end{array}\\
\begin{array}{*{20}{c}}
{{{\left[ {\frac{{{w_m}B}}{{{\nu _m}\ln 2\sum\limits_{k = n}^N {{\lambda _{m,k}}} }} - \frac{{\sigma _0^2}}{{{h_m}\left[ n \right]}}} \right]}^ + },}&{\text{otherwise,}}
\end{array}
\end{array} \right.
\end{align}
\end{subequations}
where $\lambda _{m,n}\geq0$ is the dual variable associated with the constraint $C2$; ${\left[ a \right]^ + } = \max \left( {a,0} \right)$ and $\max \left( {a,0} \right)$ denotes the bigger value  of $a$ and $0$.
\end{myTheo}
\begin{IEEEproof}
See Appendix A.
\end{IEEEproof}
\begin{rem}
It can be seen from Theorem 1 that users choose to offload their computation tasks only when the channel state information between users and the UAV is stronger than a threshold, namely, ${h_m}\left[ n \right] \ge \left( {\sigma _0^2{\nu _m}\ln 2\sum\limits_{k = n}^N {{\lambda _{m,k}}} } \right)/\left({w_m}B\right)$. This indicates that the user chooses to perform local computation when the horizontal distance between the user and the UAV is larger than $\frac{{{\beta _0}{w_m}B}}{{\sigma _0^2{\nu _m}\ln 2\sum\limits_{k = n}^N {{\lambda _{m,k}}} }} - {H^2}$. Moreover, it can be seen that the larger the weight is, the higher the chance for the user to chooses to offload its computation task. Furthermore, users prefers to offload their computation task when the local computation frequency is very large, namely, $f_m^{opt}\left[ n \right] \ge \sqrt {\frac{{\sigma _0^2{\nu _m}\ln 2}}{{3C{\gamma _c}B{h_m}\left[ n \right]}}}$.
\end{rem}
\begin{myTheo}
If there exists a time slot that $f_m^{opt}\left[ n \right]=0$, the equation $f_m^{opt}\left[ k \right]=0$ must hold, $0\leq k\leq n$.
\end{myTheo}
\begin{IEEEproof}
 Since $\lambda _{m,n}$ is the dual variable and $\lambda _{m,n}\geq0$,  from Theorem 1  $f_m^{opt}\left[ n \right]$ increases with $n$. Thus, if there exists a time slot $n$ so that $f_m^{opt}\left[ n \right]=0$, one must have $f_m^{opt}\left[ k \right]=0$, for $0\leq k\leq n$. Theorem 2 is proved.
\end{IEEEproof}
\begin{rem}
Theorem 2 indicates that the user CPU frequency increases with the time slot index. This means that the number of computation bits obtained by local computing increases with the time slot index. Moreover, the user CPU frequency  increases with the weight assigned to that user since more resources are allocated to the user with a higher weight.
\end{rem}

\begin{myTheo}
For a given trajectory $\mathbf{q}_u\left[ n \right]$, the optimal user offloading time can be obtained by solving the following equation.
\begin{align}\label{27}\ \notag
&{\log _2}\left( {1 + \frac{{{h_m}\left[ n \right]{z_m}\left[ n \right]}}{{\sigma _0^2{t_m}\left[ n \right]}}} \right) - \frac{{{h_m}\left[ n \right]{z_m}\left[ n \right]}}{{\ln 2\left\{ {\sigma _0^2{t_m}\left[ n \right] + {h_m}\left[ n \right]{z_m}\left[ n \right]} \right\}}} \\
&- \frac{{{\nu _m}N{\alpha _n}}}{{BT}} = 0.
\end{align}
\end{myTheo}
\begin{rem}
Theorem 3 can be readily proved based on the proof for Theorem 1.  Thus this proof is omitted for the sake of saving space. Moreover, $\left(15\right)$ can be solved by using the bisection method \cite{S. P. Boyd}.
\end{rem}

The values of the dual variables are needed in order to obtain the optimal CPU frequency, the optimal transmit power and the optimal offloading time for all users.  The subgradient method in Lemma 1 can be used to tackle this problem \cite{F. H. Zhou}.
\begin{lemma}
The subgradient method for obtaining the dual variables is given as
\begin{subequations}
\begin{align}\label{27}\
&\lambda _{m,n}\left( {l + 1} \right) = {\left[ {\lambda _{m,n}\left( l \right) - \theta \left( l \right)\Delta {\lambda _{m,n}}\left( l \right)} \right]^ + }, m\in \mathcal{M}, n\in \mathcal{N}\\
&{\alpha _n}\left( {l + 1} \right) = {\left[ {{\alpha _n}\left( l \right) - \vartheta \left( l \right)\Delta {\alpha _n}\left( l \right)} \right]^ + }, n\in \mathcal{N},
\end{align}
\end{subequations}
\end{lemma}
where $l$ denotes the iteration index; $ \theta \left( l \right)$ and $\vartheta \left( l \right)$ represent the iterative steps at the $l$th iteration. In $\left(16\right)$, $\Delta {\lambda _{m,n}}\left( l \right)$ and $\Delta {\alpha _n}\left( l \right)$ are the corresponding subgradients, given as
\begin{subequations}
\begin{align}\label{27}\ \notag
&\Delta {\lambda _{m,n}}\left( l \right) = \frac{{{\eta _0}T}}{N}\sum\limits_{k = 1}^n {h_m}\left[ k \right]{P_0}\\
&\ \ \ \ \ \ \ \ \ \ \ \ \ - \frac{T}{N}\sum\limits_{k = 1}^n {\left[ {{\gamma _c}\left(f_m^{l,opt}\left[ k \right]\right)^3 + {z_m}^{l,opt}\left[ k \right]} \right]}  ,\\
&\Delta {\alpha _n}\left( l \right) = {1 - \sum\limits_{m = 1}^M {{t_m^{l,opt}}\left[ n \right]} }, n\in \mathcal{N},
\end{align}
\end{subequations}
where $f_m^{l,opt}\left[ n \right]$, ${z_m}^{l,opt}\left[ n \right]$,  and ${t_m^{l,opt}}\left[ n \right]$ denote the optimal solutions at the $l$th iterations. According to \cite{S. P. Boyd}, the subgradient  guarantees  to converge to the optimal value with a  very small error range.
\subsection{Trajectory Optimization}
For any given CPU frequency,  transmit power, and offloading time of users, the trajectory optimization problem can be formulated as $\mathbf{P}_{3}$.
\begin{subequations}
\begin{align}\label{27}\ \notag
&  \mathbf{P}_{3}:\ {\mathop {\max }\limits_{\mathbf{q}_u\left[ n \right]} }\ {\sum\limits_{m = 1}^M {{w_m}} }\\
  &\times\left[ {\sum\limits_{n = 1}^N  \frac{{BT{t_m}\left[ n \right]}}{{{\nu _m}N}}{{\log }_2}\left( {1 + \frac{{{\beta _0}{P_m}\left[ n \right]}}{{\sigma _0^2}\left({H^2} + {{\left\| {{\mathbf{q}_u}\left[ n \right] - {\mathbf{q}_m}} \right\|}^2}\right)}} \right)} \right]\\ \notag
&\text{s.t.}\  C2:\frac{T}{N}\sum\limits_{k = 1}^n {\left[ {{\gamma _c}f_m^3\left[ k \right] + {t_m}\left[ k \right]{P_m}\left[ k \right]} \right]} \\
&\ \le \frac{{{\eta _0}T}}{N}\sum\limits_{k = 1}^n \frac{{{\beta _0}{P_0}}}{ {H^2} + {{\left\| {{\mathbf{q}_u}\left[ k \right] - {\mathbf{q}_m}} \right\|}^2}},m\in {\cal M},n\in {\cal N}\\
&C4 \ \text{and}\ C5.
\end{align}
\end{subequations}
Since $C2$ is non-convex and the objective function is non-concave with respect to $\mathbf{q}_u\left[ n \right]$,  $\mathbf{P}_{3}$ is non-convex and we use the SCA technique to solve the optimization problem.  The obtained solutions can be guaranteed to satisfy the Karush-Kuhn-Tucker (KKT) conditions of $\mathbf{P}_{3}$ \cite{Y. Zeng}. By using the SCA technique, Theorem 4 is given as follows.
\begin{myTheo}
For any local trajectory {$\mathbf{q}_{u,{\jmath}}\left[ n \right], n\in {\cal N}$ at the $\jmath$th iteration}, one has
\begin{subequations}
\begin{align}\label{27}\
&\sum\limits_{i = 1}^n {\frac{{ {P_0}{\beta _0}}}{{{H^2} + {{\left\| {{\mathbf{q}_u}\left[ i \right] - {\mathbf{q}_m}} \right\|}^2}}}}  \ge  {P_0}{\beta _0}\overline {{h_m}} \left[ n \right],\\
&{\overline {{h_m}} \left[ n \right] = \sum\limits_{i = 1}^n\left\{ {\frac{{{H^2} + 2{{\left\| {\mathbf{q}_{u,{\jmath}}\left[ i \right] - {\mathbf{q}_m}} \right\|}^2} - {{\left\| {{\mathbf{q}_u}\left[ i \right] - {\mathbf{q}_m}} \right\|}^2}}}{{{{\left( {{H^2} + {{\left\| {\mathbf{q}_{u,{\jmath}}\left[ i \right] - {\mathbf{q}_m}} \right\|}^2}} \right)}^2}}}}\right\}}
\end{align}
\end{subequations}
where the equality holds when {$\mathbf{q}_u\left[ n \right]=\mathbf{q}_{u,{\jmath}}\left[ n \right] $}.
\end{myTheo}
\begin{IEEEproof}
Let $f\left( z \right) = \frac{a}{{b + z}}$, where $a$ and $b$ are positive constants, and $z\geq0$. Since $f\left( z \right)$ is convex with respect to $z$, the following inequality can be obtained:
\begin{align}\label{27}\
\frac{a}{{b + z}} \ge \frac{a}{{b + {z_0}}} - \frac{a}{{{{\left( {b + {z_0}} \right)}^2}}}\left( {z - {z_0}} \right),
\end{align}
where $z_0$ is a given local point. By using $\left(20\right)$, Theorem 4 is proved.
\end{IEEEproof}

In order to tackle the objective function of $\mathbf{P}_{3}$, Lemma 2 is given as follows.
\begin{lemma} \cite{Y. Zeng}
Using the SCA method, the following inequality can be obtained,
\begin{subequations}
\begin{align}\label{27}\
&{{\log _2}\left( {1 + \frac{{{\beta _0}{P_m}\left[ n \right]}}{{\sigma _0^2\left( {{H^2} + {{\left\| {{\mathbf{q}_u}\left[ n \right] - {\mathbf{q}_m}} \right\|}^2}} \right)}}} \right) \ge {y_{m,{\jmath}}}\left( {\left\{ {{\mathbf{q}_u}\left[ n \right]} \right\}} \right),}\\ \notag
&{{y_{m,{\jmath}}}\left( {\left\{ {{\mathbf{q}_u}\left[ n \right]} \right\}} \right) = {\log _2}\left( {1 + \frac{{{\beta _0}{P_m}\left[ n \right]}}{{\sigma _0^2\left( {{H^2} + {{\left\| {\mathbf{q}_{u,{\jmath}}\left[ n \right] - {\mathbf{q}_m}} \right\|}^2}} \right)}}} \right)}\\ \notag
& - \frac{{{\beta _0}{P_m}\left[ n \right]{{\log }_2}e}}{{\left( {\sigma _0^2{H^2} + {\beta _0}{P_m}\left[ n \right] + \sigma _0^2{{\left\| {\mathbf{q}_{u,{\jmath}}\left[ n \right]} \right\|}^2}} \right)\left( {{H^2} + {{\left\| {\mathbf{q}_{u,{\jmath}}\left[ n \right]} \right\|}^2}} \right)}}\\
& \times\left( {{{\left\| {{\mathbf{q}_u}\left[ n \right]} \right\|}^2} - {{\left\| {\mathbf{q}_{u,{\jmath}}\left[ n \right]} \right\|}^2}} \right),
\end{align}
\end{subequations}
where the equality holds when {$\mathbf{q}_u\left[ n \right]=\mathbf{q}_{u,{\jmath}}\left[ n \right] $}.
\end{lemma}

\begin{table}[htbp]
\begin{center}
\caption{Two-stage alternative optimization algorithm}
\begin{tabular}{lcl}
\\\toprule
$\textbf{Algorithm 1}$: The two-stage alternative optimization algorithm\\ \midrule
\  1: \textbf{Setting:}\\
\ \  \ $P_0$, $T$, $N$, ${V_{\max }}$, $\mathbf{q}_0$, $\mathbf{q}_F$, and the tolerance errors $\xi$,  $\xi_1$; \\
\  2: \textbf{Initialization:}\\
\ \  \ The iterative number $i=1$, $\lambda _{m,n}^i$, $\alpha _n^i$ and $\mathbf{q}_u^i\left[ n \right]$; \\
\  3: \textbf{Repeat 1:}\\
 \   \ \ \ \ \ calculate $f_m^{opt,i}\left[ n \right]$ and $P_m^{opt,i}\left[ n \right]$  using Theorem 1 \\
 \   \ \ \ \ \  for given  $\mathbf{q}_u^i\left[ n \right]$; \\
 \   \ \ \ \ \ use the bisection method to solve $\left(20\right)$ and obtain ${t_m^{i,opt}}\left[ n \right]$; \\
\ \ \ \ \ \ update  $\lambda _{m,n}^i$ and $\alpha _n^i$ using the subgradient algorithm; \\
 \ \ \ \ \ \ initialize the iterative number $j=1$; \\
 \ \ \ \ \ \ \textbf{Repeat 2:}\\
  \ \ \ \ \ \ \ \ solve $\text{P}_{\textbf{4}}$ by using \texttt{CVX} for the given $f_m^{opt,i}\left[ n \right]$, $P_m^{opt,i}\left[ n \right]$\\
   \ \ \ \ \ \ \ \ and ${t_m^{i,opt}}\left[ n \right]$;\\
  \ \ \ \ \ \ \ \ update $j=j+1$, and $\mathbf{q}_u^j\left[ n \right]$;\\
 \ \ \ \ \ \ \ \  if {$\sum\limits_{n = 1}^N {\left\| {\mathbf{q}_u^j\left[ n \right] - \mathbf{q}_u^{j-1}\left[ n \right]} \right\|}  \le \xi$}   \\
  \ \ \ \ \ \ \ \  \ \ \ $\mathbf{q}_u^i\left[ n \right]=\mathbf{q}_u^j\left[ n \right]$ ;\\
   \ \ \ \ \ \ \ \  \ \ \ break; \\
 \ \ \ \ \ \ \ \  end \\
  \ \ \ \ \ \  \textbf{end Repeat 2} \\
 \ \ \ \ \ update the iterative number $i=i+1$;  \\
\ \ \ \ \ if $\left|R^{i}-R^{i-1}\right|\leq \xi_1$ \\
\ \ \ \ \  \  break;\\
\ \ \ \ \  end\\
\ \ \ \ \textbf{end Repeat 1}\\
\  4: \textbf{Obtain solutions:}\\
 \ \ \ \ \ \ $f_m^{opt}\left[ n \right]$, $P_m^{opt}\left[ n \right]$ and ${t_m^{opt}}\left[ n \right]$ and $\mathbf{q}_u^{opt}\left[ n \right]$. \\
\bottomrule
\end{tabular}
\end{center}
\end{table}

Using Theorem 4 and Lemma 2, $\mathbf{P}_{3}$ can be solved by iteratively solving the approximate problem $\mathbf{P}_{4}$, given as
\begin{subequations}
\begin{align}\label{27}\
&{\mathbf{P}_{4}:\ {\mathop {\max }\limits_{\mathbf{q}_u\left[ n \right]} }\ {\sum\limits_{m = 1}^M {{w_m}\left[ {\sum\limits_{n = 1}^N  \frac{{BT{t_m}\left[ n \right]{y_{m,{\jmath}}}\left( {\left\{ {{\mathbf{q}_u}\left[ n \right]} \right\}} \right)}}{{{\nu _m}N}}} \right]} }}\\
&\text{s.t.} \ \ \  C4 \ \text{and}\ C5, \\ \notag
& \sum\limits_{k = 1}^n {\left[ {{\gamma _c}f_m^3\left[ k \right] + {t_m}\left[ k \right]{P_m}\left[ k \right]} \right]}  \le{{{\eta _0}{P_0}{\beta _0}\overline {{h_m}} \left[ n \right]}},\\
&\ \ \ \ \ \ \ \ \ \ \ \ \ \ \ \ \ \ \ \ \ \ \ \ \ \ \ \ m\in {\cal M},n\in {\cal N}.
\end{align}
\end{subequations}
It can be seen that $\mathbf{P}_{4}$ is convex and can be readily solved by using  \texttt{CVX} \cite{F. Zhou2}.  By solving $\mathbf{P}_{2}$ and $\mathbf{P}_{4}$, a two-stage alternative optimization algorithm denoted by Algorithm 1 is further developed  to solve $\mathbf{P}_{1}$. The details for Algorithm 1 can be found in Table I. In Table I,  $R^{i}$ denotes the value of the objective function of $\mathbf{P}_{{1}}$ at the $i$th iteration.
\section{Resource Allocation in Binary Computation Offloading Mode}
In this section, the weighted sum computation bits maximization problem is studied in the UAV-enabled wireless powered MEC system under the binary computation offloading mode. The CPU frequencies of the users that choose to perform local computation, the offloading times,  the transmit powers of users that  choose to perform task offloading, the trajectory of the UAV,  and the mode selection are jointly optimized to maximize the weighted sum computation bits of all users. The formulated problem is a mixed integer non-convex optimization problem, for which a three-stage alternative optimization problem is proposed.
\subsection{Resource Allocation Problem Formulation}
Under  the binary computation offloading mode,  the weighted sum computation bit maximization problem subject to the energy harvesting causal constraints, the UAV  speed and position constraints is formulated as $\mathbf{P}_{5}$,
\begin{subequations}
\begin{align}\label{27}\ \notag
 & \mathbf{P}_{5}:{\mathop {\max }\limits_{\scriptstyle{f_i}\left[ n \right],{P_j}\left[ n \right],q\left[ n \right],\hfill\atop
\scriptstyle{t_j}\left[ n \right], {{\cal M}_0}, {{\cal M}_1}\hfill}
 }\ {\sum\limits_{i \in {{\cal M}_0}} {\sum\limits_{n = 1}^N {{w_i}\frac{{{f_i}\left[ n \right]T}}{{CN}}} }   }\\
 & + \sum\limits_{j \in {{\cal M}_1}} {\frac{{{w_j}BT}}{{{\nu _j}N}}\sum\limits_{n = 1}^N {{t_j}\left[ n \right]{{\log }_2}} \left( {1 + \frac{{{h_j}\left[ n \right]{P_j}\left[ n \right]}}{{\sigma _0^2}}} \right)}\\
 &\text{s.t.}\ \frac{T}{N}\sum\limits_{k = 1}^n {{\gamma _c}f_i^3\left[ k \right]}  \le \frac{{{\eta _0}T}}{N}\sum\limits_{k = 1}^n {{h_i}\left[ k \right]{P_0}} , n\in {\cal N}, i\in {{\cal M}_0},\\
& \frac{T}{N}\sum\limits_{k = 1}^n {{t_j}\left[ k \right]{P_j}\left[ k \right]}  \le \frac{{{\eta _0}T}}{N}\sum\limits_{k = 1}^n {{h_j}\left[ k \right]{P_0}} , n\in {\cal N}, j\in {{\cal M}_1},\\
& \sum\limits_{j \in {{\cal M}_1}} {{t_j}\left[ n \right]}  \le 1, n\in {\cal N},\\
&  {\cal M}={\cal M}_0\cup{\cal M}_1, {\cal M}_0 \cap{\cal M}_1=\Theta,\\
&  {f_i}\left[ n \right]\geq0,  {P_j}\left[ n \right]\geq0, i\in {{\cal M}_0}, j\in {{\cal M}_1},\\
& C4 \ \text{and}\ C5.
\end{align}
\end{subequations}
$\left(23\rm{b}\right)$ and $\left(23\rm{c}\right)$ are the energy harvesting causal constraints imposed on these users who choose to perform local computation and on these users who choose to perform task offloading, respectively; $\left(23\rm{d}\right)$ is the offloading time constraint during each slot and $\left(23\rm{e}\right)$ is the user operation selection constraint. In $\mathbf{P}_{{5}}$ there exist close couplings among different optimization variables. Furthermore, the binary user operation mode selection  makes $\mathbf{P}_{{5}}$ a mixed integer programming problem.  The exhaustive search method leads to a  prohibitively  high computational complexity,  especially when there exist a  large number  of users. Motivated by how we solve $\mathbf{P}_{{1}}$, $\mathbf{P}_{{5}}$ has a similar structure as $\mathbf{P}_{{1}}$ when the operation modes of users are determined. Thus, the optimal CPU frequency, transmit power, and offloading time of users can be obtained by using the same method as the one  used for $\text{P}_{{1}}$ and the trajectory optimization for the UAV can also be achieved by using the SCA method. As such, a three-stage alternative optimization algorithm is proposed based on the two-stage Algorithm 1.  The details for the algorithm are presented as follows.
\subsection{Three-Stage Alternative Optimization Algorithm}
In order to efficiently solve $\mathbf{P}_{{5}}$, a binary variable denoted by ${{\rho _m}}$ is introduced, where ${\rho _m} \in \left\{ {0,1} \right\}$ and $m\in {{\cal M}}$. ${{\rho _m}}=0$ indicates that the $m$th user performs local computation mode while ${{\rho _m}}=1$ means that the $m$th user  performs task offloading. Moreover, the user operation selection indicator variable $\rho _{m}$ is relaxed as a sharing factor $\rho _{m}\in \left[0, 1\right]$. Thus, $\mathbf{P}_{{5}}$ can be rewritten as
\begin{subequations}
\begin{align}\label{27}\ \notag
& \mathbf{P}_{{6}}: {{\mathop {\max }\limits_{\scriptstyle{f_m}\left[ n \right],{P_n}\left[ n \right],\mathbf{q}\left[ n \right],\hfill\atop
\scriptstyle{t_m}\left[ n \right],{\rho _m}\hfill} }
 }\ \sum\limits_{m = 1}^M \sum\limits_{n = 1}^N {w_m}\left\{ \left( {1 - {\rho _m}} \right)\frac{{{f_m}\left[ n \right]T}}{{CN}} \right. \\
 &\ \ \ \left.+ \frac{{BT{t_m}\left[ n \right]{\rho _m}}}{{{\nu _m}N}}{{\log }_2}\left( {1 + \frac{{{h_m}\left[ n \right]{P_m}\left[ n \right]}}{{\sigma _0^2}}} \right) \right\} \\ \notag
 & \text{s.t.}\ \left( {1 - {\rho _m}} \right)\frac{T}{N}\sum\limits_{k = 1}^n {{\gamma _c}f_m^3\left[ k \right]}  + {\rho _m}\frac{T}{N}\sum\limits_{k = 1}^n {{t_m}\left[ k \right]{P_m}\left[ k \right]}  \\
&\le \frac{{{\eta _0}T}}{N}\sum\limits_{k = 1}^n {{h_m}\left[ k \right]{P_0}},m\in {{\cal M}},\\
& \sum\limits_{m = 1}^M {{\rho _m}{t_m}\left[ n \right]}  \le 1, n\in {\cal N},\\
& {f_m}\left[ n \right]\geq0,  {P_m}\left[ n \right]\geq0, n\in {\cal N}, m\in {{\cal M}},\\
& C4 \ \text{and}\ C5.
\end{align}
\end{subequations}
Even by relaxing the binary variable ${{\rho _m}}$, $\mathbf{P}_{{6}}$ is still difficult to solve as there exist couplings among different variables. For any given $\rho _{m}$ and the trajectory of the UAV, $\mathbf{P}_{{6}}$ has a similar structure as $\mathbf{P}_{{1}}$. Thus, using the same techniques applied to $\mathbf{P}_{{1}}$, the optimal CPU frequency, transmit power and offloading time of users for a given $\rho _{m}$ and the UAV  trajectory  can be obtained. It is easy to verify that the optimal CPU frequency, transmit power and offloading time of users for a given trajectory have the same forms given by Theorem 1 and Theorem 3.

\begin{myTheo}
For any given $f_m\left[ n \right]$, ${P_m}\left[ n \right]$, $t_m\left[ n \right]$ and $\mathbf{q}_u\left[ n \right]$, the user operation selection scheme can be obtained by
\begin{subequations}
\begin{align}\label{27}\
&{\rho _m^{opt}}{\rm{ = }}\left\{ \begin{array}{l}
\begin{array}{*{20}{c}}
{0}&{\text{if}\ {{G}}_1 \ge {{G}_2},}
\end{array}\\
\begin{array}{*{20}{c}}
{1}&{\text{otherwise;}}
\end{array}
\end{array} \right.\\
&{G}_1 = \sum\limits_{n = 1}^N {\left\{ {\frac{{{w_m}{f_m}\left[ n \right]}}{C} - {\upsilon _{m,n}}\sum\limits_{k = 1}^n {{\gamma _c}f_m^3\left[ k \right]} } \right\}},\\ \notag
&{G}_2 = \sum\limits_{n = 1}^N \left\{ \frac{{B{t_m}\left[ n \right]}}{{{\nu _m}}}{{\log }_2}\left( {1 + \frac{{{h_m}\left[ n \right]{P_m}\left[ n \right]}}{{\sigma _0^2}}} \right)\right. \\
&\left.- {\upsilon _{m,n}}\sum\limits_{k = 1}^n {{t_m}\left[ k \right]{P_m}\left[ k \right]}  - \frac{N}{T}{\varepsilon _n}{t_m}\left[ n \right] \right\},
\end{align}
\end{subequations}
where ${\upsilon _{m,n}}\geq0$ and ${\varepsilon _n}\geq0$ are the dual variables associated with the constraints given by $\left(24\rm{b}\right)$ and $\left(24\rm{c}\right)$, respectively.
\end{myTheo}
\begin{IEEEproof}
See Appendix B.
\end{IEEEproof}
\begin{rem}
Theorem 5 indicates that the user operation selection scheme depends on the tradeoff between the achievable computation rate and the operation cost. If the tradeoff of the user achieved by local computing is better than that obtained by task offloading, the user chooses to perform local computing; otherwise, the user chooses to offload its computation tasks to the UAV for computing.
\end{rem}

Finally, the trajectory optimization for any given $\rho _{m}$, $f_m\left[ n \right]$, ${P_m}\left[ n \right]$ and $t_m\left[ n \right]$ can be obtained by solving $\mathbf{P}_{{7}}$, given as
\begin{subequations}
\begin{align}\label{27}\
&{\mathbf{P}_{{7}}:\ {\mathop {\max }\limits_{\mathbf{q}_u\left[ n \right]} }\ {\sum\limits_{m = 1}^M {{w_m}\rho _{m}\left[ {\sum\limits_{n = 1}^N  \frac{{BT{t_m}\left[ n \right]{y_{m,{\jmath}}}\left( {\left\{ {{\mathbf{q}_u}\left[ n \right]} \right\}} \right)}}{{{\nu _m}N}}} \right]} }}\\
&\text{s.t.} \ \ \  C4 \ \text{and}\ C5,\\ \notag
& \left( {1 - {\rho _m}} \right)\sum\limits_{k = 1}^n {{\gamma _c}f_m^3\left[ k \right]}  + {\rho _m}\sum\limits_{k = 1}^n {{t_m}\left[ k \right]{P_m}\left[ k \right]}  \\
&\le{{{\eta _0}{P_0}{\beta _0}\overline {{h_m}} \left[ n \right]}},m\in {\cal M},n\in {\cal N},
\end{align}
\end{subequations}
where $\overline {{h_m}} \left[ n \right]$ and ${y_j}\left( {\left\{ {{\mathbf{q}_u}\left[ n \right]} \right\}} \right)$ are given by $\left(19\rm{b}\right)$ and $\left(21\rm{b}\right)$, respectively.   $\mathbf{P}_{{7}}$ is convex and can be efficiently solved by using  \texttt{CVX} \cite{F. Zhou2}. Based on Theorem 1, Theorem 5 and the solutions of $\mathbf{P}_{{7}}$, a three-stage alternative optimization algorithm denoted by Algorithm 2 is proposed to solve $\mathbf{P}_{{5}}$. The details for Algorithm 2 are presented in Table 2. In Table 2,  $R^{l}$ and $R^{i}$ denote the value of the objective function of $\mathbf{P}_{{5}}$ at the $l$th and $i$ iteration, respectively.
\begin{table}[htbp]
\begin{center}
\caption{Three-stage alternative optimization algorithm}
\begin{tabular}{lcl}
\\\toprule
$\textbf{Algorithm 2}$: The three-stage alternative optimization algorithm \\ \midrule
\  1: \textbf{Setting:}\\
\ \  \ $P_0$, $T$, $N$, ${V_{\max }}$, $\mathbf{q}_0$, $\mathbf{q}_F$, and the tolerance errors $\xi$,  $\xi_1$ and $\xi_2$; \\
\  2: \textbf{Initialization:}\\
\ \  \ The iterative number $i=1$, ${\upsilon _{m,n}^i}$ and ${\varepsilon _n^i}$, and $\mathbf{q}_u^i\left[ n \right]$; \\
\  3: \textbf{Repeat 1:}\\
\ \ \ \ \ \ initialize the iterative number $l=1$ and $\rho _{m}^l$; \\
\  \ \ \ \ \ \ \textbf{Repeat 2:}\\
 \   \ \ \ \ \ \ \ calculate $f_m^{opt,i}\left[ n \right]$ and $P_m^{opt,i}\left[ n \right]$  using Theorem 1  \\
 \   \ \ \ \ \ \ \ for given  $\mathbf{q}_u^i\left[ n \right]$ and $\rho_m^{opt,l}$; \\
 \   \ \ \ \ \ \ \ use the bisection method to solve $\left(20\right)$ and obtain ${t_m^{i,opt}}\left[ n \right]$; \\
\ \ \ \ \ \ \ \ update  ${\upsilon _{m,n}^i}$ and ${\varepsilon _n^i}$ using the subgradient algorithm; \\
\ \ \ \ \ \ \ \ calculate  $\rho_m^{opt,l}$ using Theorem 5 and update $l=l+1$; \\
\ \ \ \ \ \ \ \ if $\left|R^{l}-R^{l-1}\right|\leq \xi$ \\
\ \ \ \ \  \ \ \ \   break;\\
\ \ \ \ \ \ \ \  end\\
 \ \ \ \ \ \ \ \ initialize the iterative number $j=1$; \\
 \ \ \ \ \ \ \textbf{Repeat 3:}\\
  \ \ \ \ \ \ \ \ solve $\text{P}_{\textbf{7}}$ by using \texttt{CVX} for the given $f_m^{opt,i}\left[ n \right]$, $P_m^{opt,i}\left[ n \right]$, \\
   \ \ \ \ \ \ \ \  ${t_m^{i,opt}}\left[ n \right]$ and $\rho_m^{opt,l}$;\\
  \ \ \ \ \ \ \ \ update $j=j+1$, and $\mathbf{q}_u^j\left[ n \right]$;\\
 \ \ \ \ \ \ \ \  if $\sum\limits_{n = 1}^N {\left\| {\mathbf{q}_u^j\left[ n \right] - \mathbf{q}_u^{j-1}\left[ n \right]} \right\|}  \le \xi$   \\
  \ \ \ \ \ \ \ \  \ \ \ $\mathbf{q}_u^i\left[ n \right]=\mathbf{q}_u^j\left[ n \right]$ ;\\
   \ \ \ \ \ \ \ \  \ \ \ break; \\
 \ \ \ \ \ \ \ \  end \\
  \ \ \ \ \ \  \textbf{end Repeat 3} \\
 \ \ \ \ \ update the iterative number $i=i+1$;  \\
\ \ \ \ \ if $\left|R^{i}-R^{i-1}\right|\leq \xi_1$ \\
\ \ \ \ \  \  break;\\
\ \ \ \ \  end\\
\  \ \ \ \  \textbf{ end Repeat 2}\\
\ \ \ \ \textbf{end Repeat 1}\\
\  4: \textbf{Obtain solutions:}\\
 \ \ \ \ \ \ $f_m^{opt}\left[ n \right]$, $P_m^{opt}\left[ n \right]$ and ${t_m^{opt}}\left[ n \right]$, $\rho_m^{opt}$ and $\mathbf{q}_u^{opt}\left[ n \right]$. \\
\bottomrule
\end{tabular}
\end{center}
\end{table}

\subsection{Complexity Analysis}
The complexity of Algorithm 1 comes from four aspects. The first aspect is from the computation of the CPU frequency and the offloading power. The second aspect is from the bisection method for obtaining the offloading time. The third aspect is from the subgradient method for computing the dual variables. The fourth aspect comes from the application of \texttt{CVX} for solving $\text{P}_{\textbf{4}}$. Let $L_1$ and $L_2$ denote the number of iterations required for the outer loop and the inner loop of Algorithm 1, respectively. Let $\ell_1$ and $\ell_2$ denote the tolerance error for the bisection method and the subgradient method, respectively. Thus, according to the works in \cite{S. P. Boyd}, \cite{S. Bubeck} and \cite{C. Gutierrez}, the total complexity of Algorithm 1 is $
{\cal O}\left[{L_1}\left(2MN+M\log_2{\left(\ell_1/T\right)+1/\ell_2^2 +L_2N^3 }\right) \right]$ and ${\cal O}\left( {\cdot} \right)$ is the big-$\rm O$ notation \cite{S. P. Boyd}.

The complexity of Algorithm 2 comes from five aspects. Four aspects are the same as these of Algorithm 1. The fifth aspect is from the computation of the operation selection indicator variable $\rho _{m}$. Let $L_1$, $L_2$ and $L_3$ denote the number of iterations required for the first, second and third loop of Algorithm 2, respectively.  Similar to the complexity analysis for Algorithm 1, the total complexity of Algorithm 2 is $
{\cal O}\left[{L_1}{L_2}\left(2MN+M+M\log_2{\left(\ell_1/T\right)+1/\ell_2^2 +L_3N^3 }\right) \right]$.

\section{Simulation Results}
In this section, simulation results are presented to compare the performance of our proposed designs  with that of other benchmark schemes. The convergence performance of the proposed algorithms is also evaluated. The simulation settings are based on the works in  \cite{C. Wang}, \cite{F. Wang}, \cite{S. Bi} and \cite{S. Jeong}. The positions of users are set as: $\mathbf{q}_1=[0,0]$, $\mathbf{q}_2=[0,10]$, $\mathbf{q}_3=[10,10]$, $\mathbf{q}_4=[10,0]$.  The detailed settings are given in Table III. The weight vector of each user $\left[w_1\ w_2\ w_3\ w_4\right]$ is set as $[0.1 \ 0.4 \ 0.3\  0.2]$.
\begin{table}[htbp]
\centering
 \caption{\label{tab:test}Simulation Parameters}
 \begin{tabular}{l|c|c}
  \midrule
  \midrule
  Parameters & Notation & Typical Values  \\
  \midrule
  \midrule
 Numbers of Users & $M$ & $4$ \\
The height of the UAV & $H$ & $10$ m \\
The time length of the UAV flying & $T$ & $2$ sec \\
Numbers of  CPU cycles & $C$ & $10^3$ cycles/bit \\
Energy conversation efficiency & $\eta_0$ & $0.8$ \\
Communication bandwidth & $B$ & $40$ MHz \\
The receiver noise power & $\sigma_0^2$ & $10^{-9}$ W\\
The number of time slots & $N$ & $50$  \\
The effective switched capacitance & $\gamma_c$ & $10^{-28}$ \\
The channel power gain &$\beta_0$& $-50$ dB\\
 The tolerance error & $\xi, \xi_1$ & $10^{-4}$ \\
 The initial position of the UAV & $\mathbf{q}_0$ & $[0, 0]$ \\
  The final position of the UAV & $\mathbf{q}_F$ & $[10, 0]$ \\
The maximum speed of the UAV & $V_{\max }$  & $20$ m/s\\
\midrule
 \end{tabular}
\end{table}

\begin{figure}[!t]
\centering
\includegraphics[width=3.0 in]{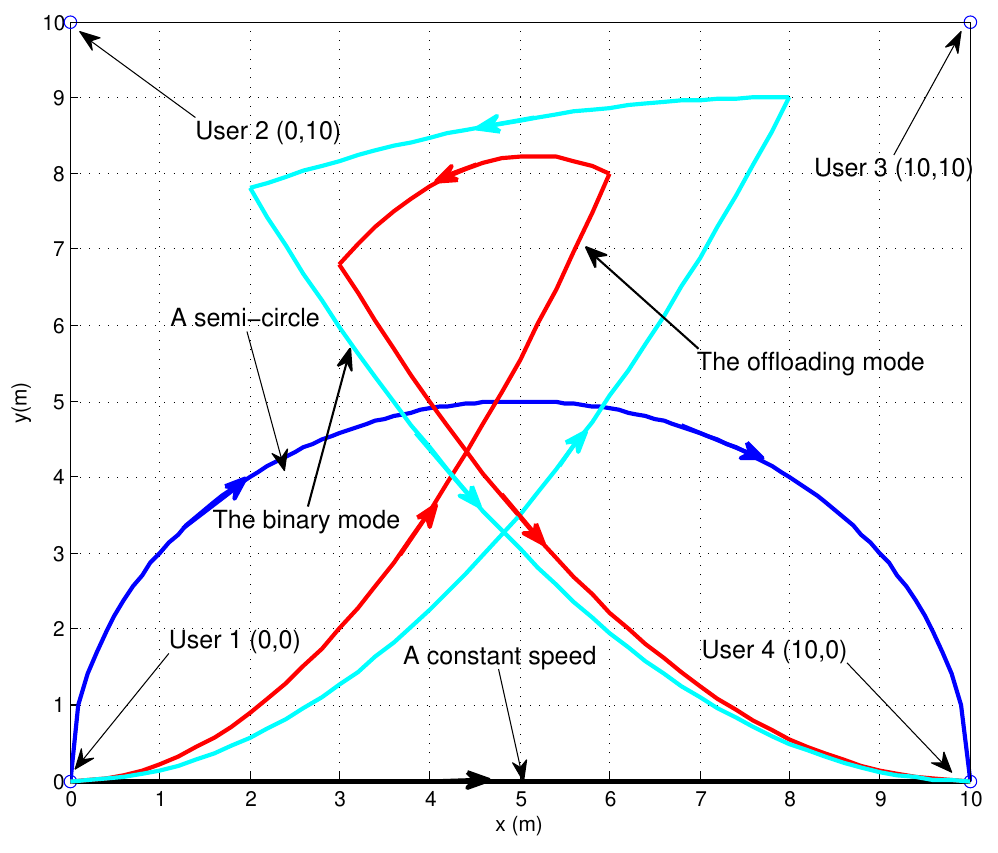}
\caption{The trajectory of the UAV under different schemes with $T=2$ seconds.} \label{fig.1}
\end{figure}
Fig. 3 shows the UAV trajectory under different schemes with $T=2$ seconds. The UAV transmit power is set as $P_0=0.1$ W. In the constant speed scenario, the UAV flies straight with a constant speed from the initial position to the final position. In the semi-circle scenario, the UAV flies along the trajectory that is a semi-circle with its diameter being ${\left\| {\mathbf{q}_F - {\mathbf{q}_0}} \right\|}$. The trajectory of the offloading mode is  obtained by using Algorithm 1 for the partial computation offloading mode and the trajectory of the binary mode is obtained by using Algorithm 2 for the binary computation offloading mode. It can be seen from  the trajectories of our proposed schemes the UAV is always close to user 2 and user 3, irrespective of the operation modes. The reason is that the weights of user 2 and user 3 are larger than these of user 1 and user 4. Thus, the UAV needs to fly close to user 2 and user 3 so as to provide more energy to them. This indicates that the priority and the fairness among users can be obtained by using the weight vector.

\begin{figure}[!t]
\centering
\includegraphics[width=3.0 in]{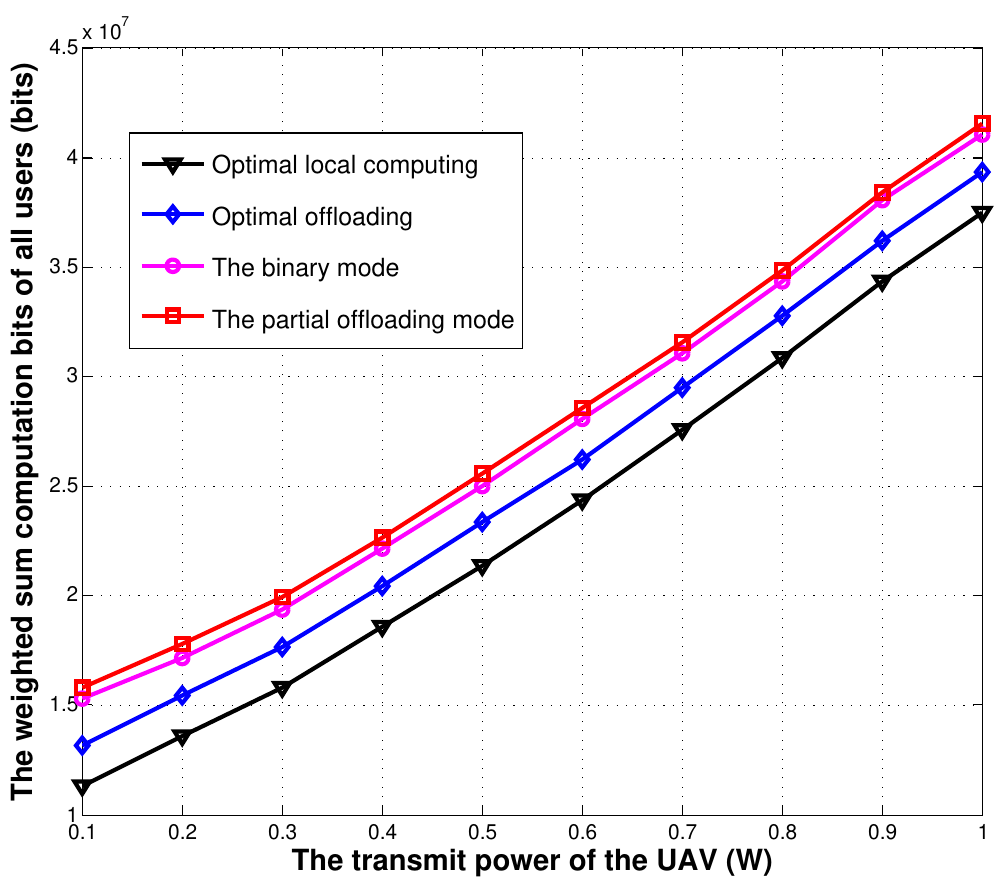}
\caption{The weighted sum computation bits of all users versus the transmit power of the UAV under different schemes.} \label{fig.1}
\end{figure}
Fig. 4 shows the weighted sum computation bits of all users versus the transmit power of the UAV under different schemes. The optimal local computing is the mode that all users only perform local computing while the optimal offloading mode is that all users only perform task offloading. And the trajectory  of the UAV is jointly optimized under these two benchmark schemes. The results  under the binary mode and the partial offloading mode are obtained by using Algorithm 2 and Algorithm 1, respectively. In Fig. 4  the weighted sum computation bits achieved under the partial offloading mode is the largest among these obtained by other schemes. The reason is that all the users can dynamically select the operation mode based on the quality of the channel state information under the partial computation offloading mode. Moreover,  the optimal offloading mode outperforms the optimal local computing. This result is consistent with the results obtained in \cite{C. You}. Furthermore, the weighted sum computation bits of all users increase with the UAV transmit power. It can be explained by the fact that the harvesting energy increases with the transmit power of the UAV. Thus, users have more energy to perform local commutating or task offloading.

\begin{figure}[htb]
\centering
\includegraphics[width=3.0 in,height=3.0 in]{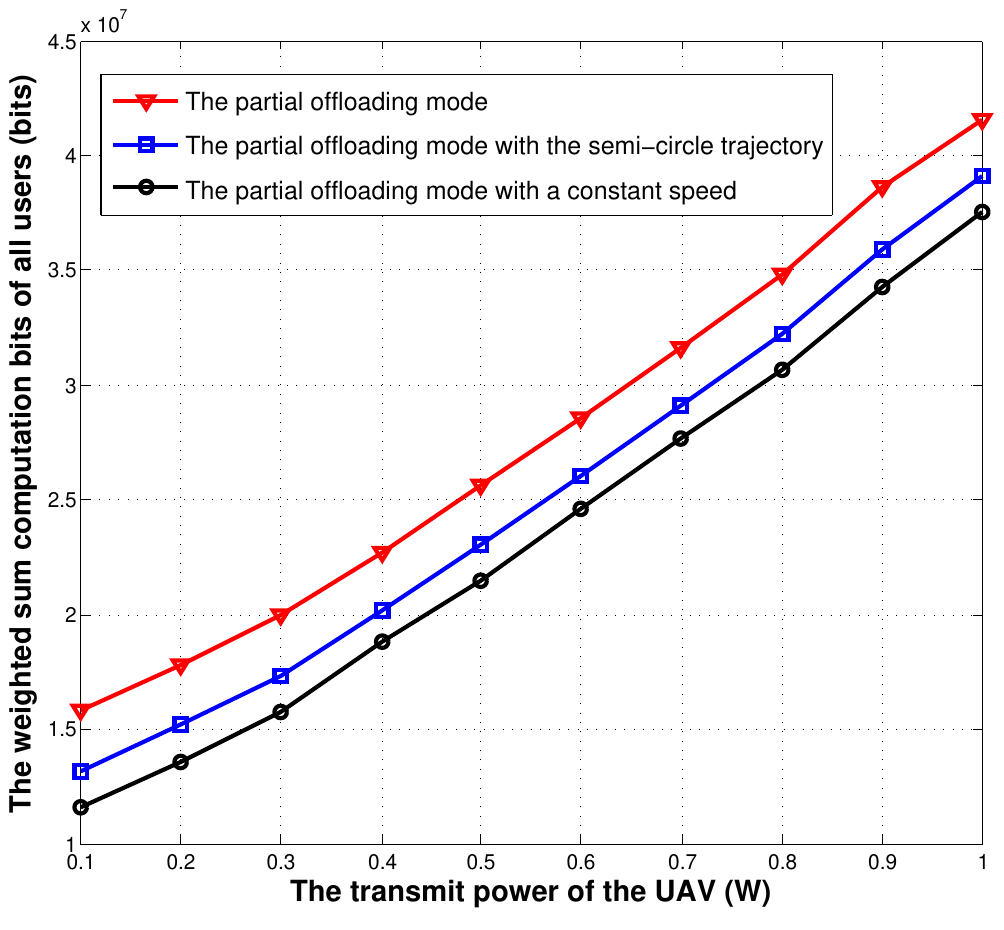}
\includegraphics[width=3.0 in,height=3.0 in]{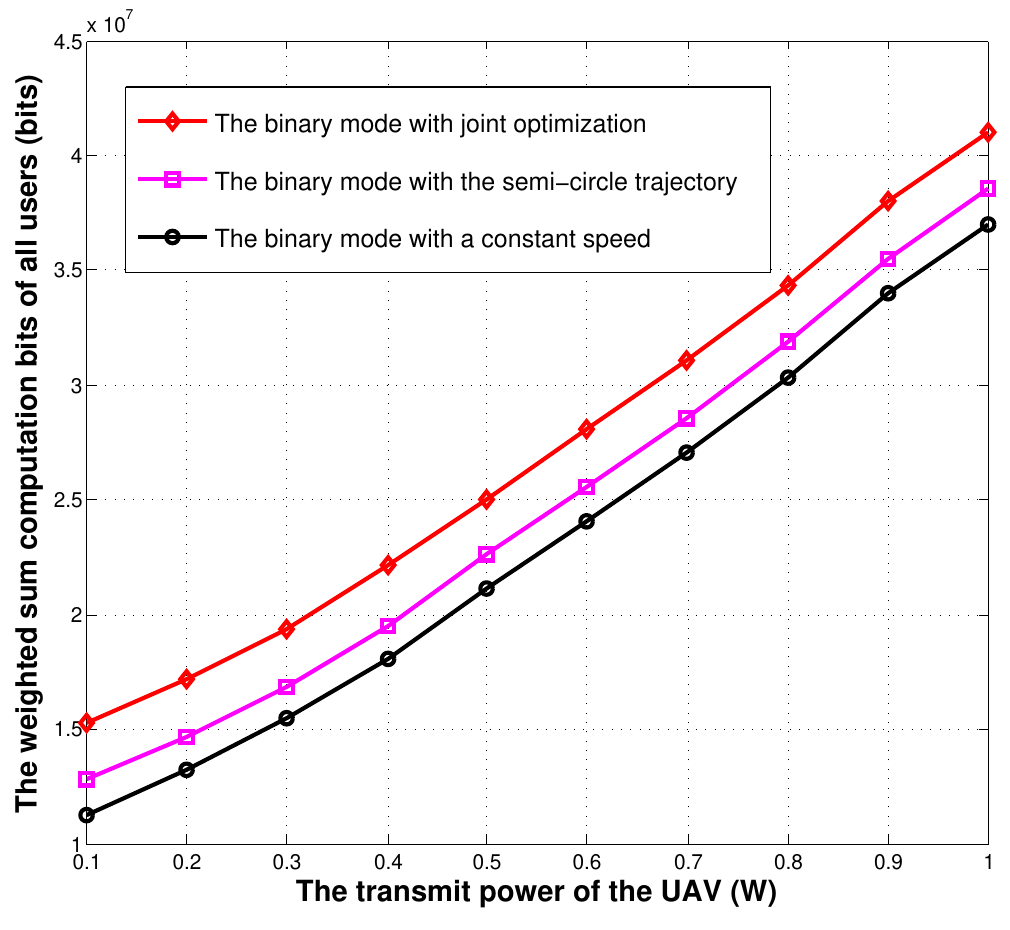}
\put(-308,-10){\footnotesize{(a)}}
\put(-100,-10){\footnotesize{(b)}}
\caption{(a) The weighted sum computation bits of all users versus the transmit power of the UAV under different  trajectories with the partial  computation offloading mode; (b) The weighted sum computation bits of all users versus the transmit power of the UAV under different  trajectories with the binary  computation offloading mode.} \label{fig1}
\end{figure}
Fig. 5 shows the weighted sum computation bits of all the users versus the transmit power of the UAV under different  trajectories with the partial  computation offloading mode and the binary  computation offloading mode. As shown in Fig. 5, the weighted sum computation bits of all the users achieved by using our proposed schemes are larger than that obtained by using the trajectory with a constant speed and than that obtained by using the semi-circle trajectory, irrespective of the operation modes. This indicates that the optimization of the trajectory of the UAV can improve the weighted sum computation bits. It also verifies that our proposed resource allocation scheme outperforms  the disjoint optimization schemes.

\begin{figure}[!t]
\centering
\includegraphics[width=3.0 in]{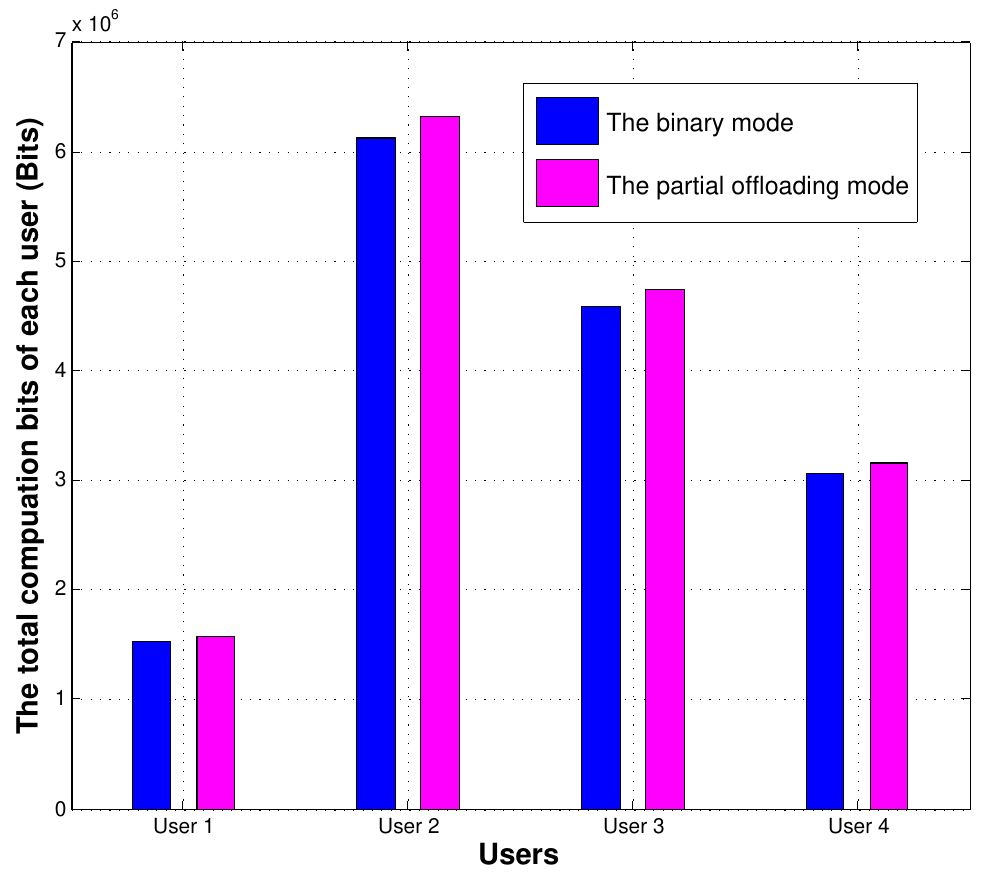}
\caption{The total computation bits of each user under different operation modes with $P_0=0.1$ W.} \label{fig.1}
\end{figure}
Fig. 6 shows the total computation bits of each user under different operation modes. The transmit power of the UAV is set as $P_0=0.1$ W. The total computation bits of user 2 and user 3 are higher than those of user 1 and user 4. The reason is that the weights of user 2 and user 3 are larger than those of user 1 and user 4. Thus, the resource allocation scheme should consider the priority of user 2 and user 3. This further verifies that the application of the weight vector can improve the priority and also the fairness of users.

\begin{figure}[!t]
\centering
\includegraphics[width=3.0 in]{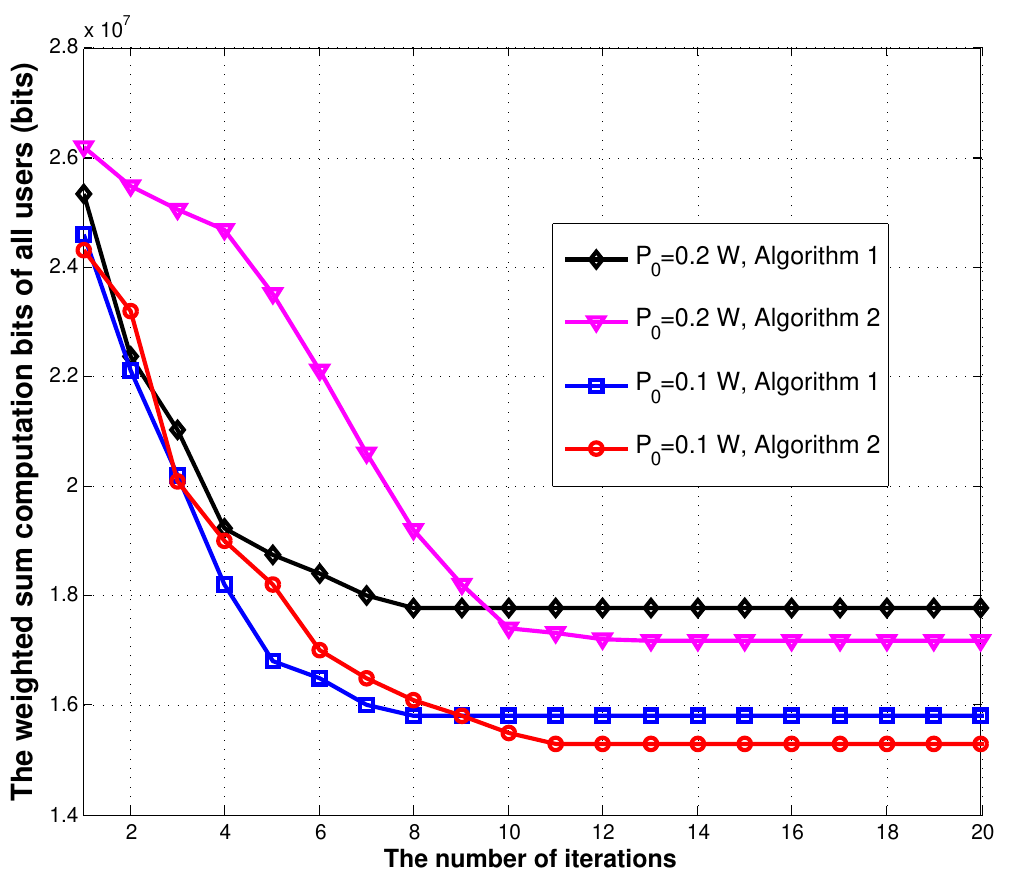}
\caption{The weighted sum computation bits of all users versus the number of iterations required by using Algorithms 1  and  2 under different transmit powers of the UAV and different operation modes.} \label{fig.1}
\end{figure}
Fig. 7 is given to verify the efficiency of our proposed Algorithm 1 and Algorithm 2. The transmit power of the UAV is  given as $0.1$ W or $0.2$ W.  The results show that  Algorithm 1 and Algorithm 2 only need several iterations to converge. This indicates that the proposed Algorithm 1 and Algorithm 2 are computationally effective  and have a fast convergence rate. It can also be seen that the weighted sum computation bits of all the  users achieved under the  partial  computation offloading mode are larger than those obtained under the binary computation offloading mode. The reason is that users can simultaneously perform local computing and  task offloading when the channel state information is strong under the partial  computation offloading mode. However, users can only perform either  local computing or task offloading in the binary offloading mode even when the channel state information is strong. The computation performance is improved by the flexible selection of the operation mode based on the channel state information.

\begin{figure}[!t]
\centering
\includegraphics[width=3.0 in]{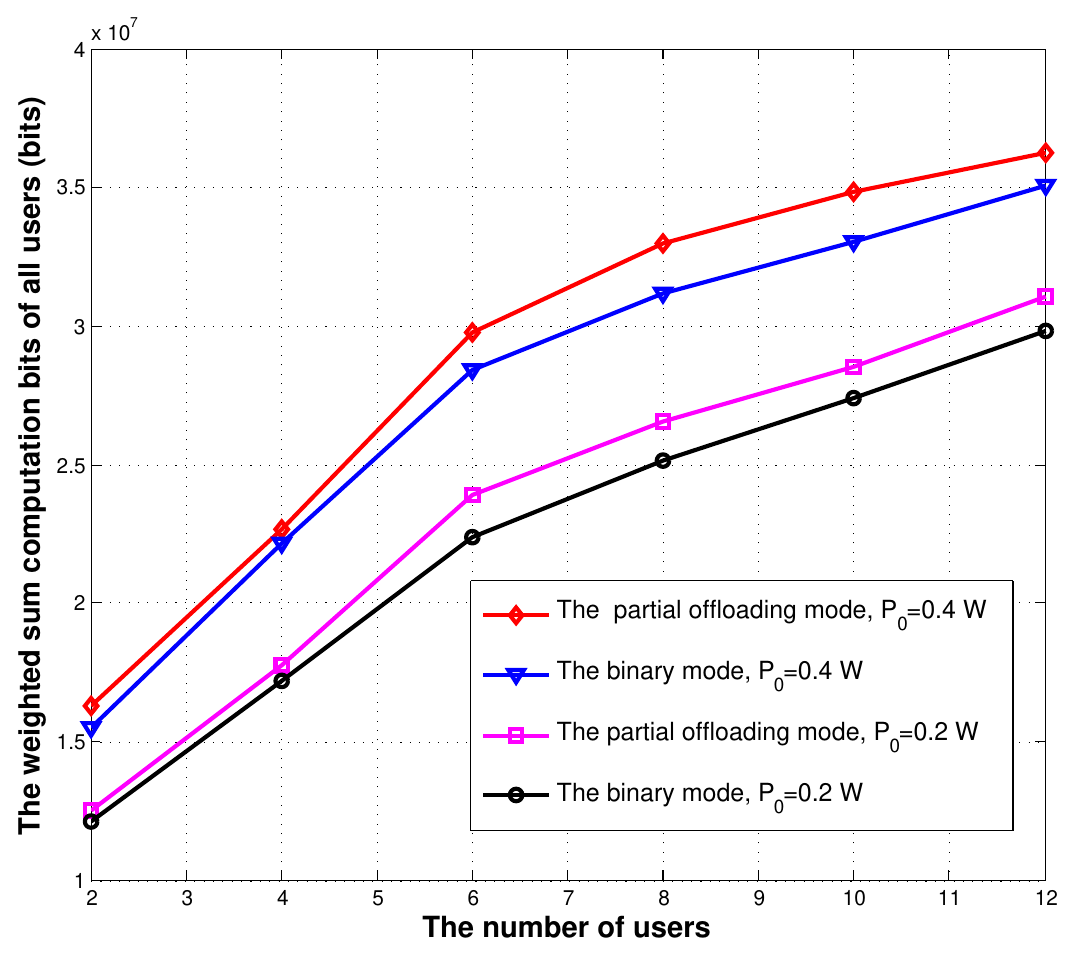}
\caption{The weighted sum computation bits of all users versus the number of users under different transmit powers of the UAV and different operation modes.} \label{fig.1}
\end{figure}
Fig. 8 shows the weighted sum computation bits of all users versus the number of users under different operation modes. The transmit power of the UAV is set as $P_0=0.2$ W or $P_0=0.4$ W.  In Fig. 8 the weighted sum computation bits of all users increase with the number of users. The reason is that more users can exploit the harvesting energy to perform local computing and computation offloading. It is also observed  that the growth rate decreases with the increase of the number of users. The reason is that the offloading time allocated for each user decreases with the increase of the number of users since the total offloading time is limited by $T$.

\begin{table*}
\begin{center}
\caption{Comparison of  the required run time of Algorithm 1 with that of Algorithm 2 (s)}
\begin{tabular}{|c|c|c|c|c|c|c|c|c|c|}\hline
\backslashbox{Algorithms}{$ \left(N, M\right)$}          &$\left(50, 2\right)$ &$\left(50, 4\right)$ &$\left(50, 8\right)$ &$\left(60, 2\right)$&$\left(60, 4\right)$&$\left(60, 8\right)$&$\left(70, 2\right)$&$\left(70, 4\right)$&$\left(70, 8\right)$\\\hline
Algorithm 1 &43.72 &104.54&186.38&154.74 &198.65&235.85&224.74&291.53&352.72\\\hline
Algorithm 2 &89.35&167.17&265.46&223.19&275.42&321.87&308.56&388.92&468.39 \\\hline
\end{tabular}
\end{center}
\end{table*}

Table IV is given to evaluate the run times of Algorithm 1 and Algorithm 2 shown in the top of the next page. The run times are obtained by using a computer with 64-bit Intel(R) Core(TM) i7-4790 CPU, $8$ GB RAM. From Table IV we can see that the required run time of Algorithm 1 is smaller than that of Algorithm 2. This indicates that the complexity of Algorithm 1 is lower than that of Algorithm 2. It can be verified by the complexity analysis presented in Subsection C of Section IV. Moreover, the effect of the number of time slots on the run time is larger than that of the number of users. The reason is that the complexity of these two algorithms mainly depends on the number of time slots. This can also be verified by the complexity analysis.

\section{Conclusions}
The resource allocation problems were studied for UAV-enabled wireless powered MEC systems under both the partial and binary computation offloading modes. The weighted sum  computation rates of users were maximized by jointly optimizing  the CPU frequencies, the user offloading times, the user  transmit powers, and the UAV  trajectory Two alternative algorithms were proposed to solve these challenging problems. The closed-form expressions for the optimal CPU frequencies, user  offloading times, and user transmit power were derived. Moreover, the optimal selection scheme whether users choose to locally compute or offload tasks was proposed for the binary computation offloading mode. It was shown that the performance achieved by using our proposed resource allocation scheme is superior to these obtained by using the disjoint optimization schemes. Simulation results also verified the efficiency of our proposed alternative algorithms and our theoretical analysis.

The exploitation of UAV to improve the energy conversation efficiency and the computation performance was studied in this paper. However, the computation performance is also limited by the flight time of the UAV. It is interesting to exploit multiple antennas techniques to tackle this challenge. This will be investigated in our future work.
\appendices
\section{Proof of Theorem 1}
Let $\lambda _{m,n}$ and ${{\alpha _n}}$ denote the dual variables associated with the constraint $C2$ and $C3$, respectively, where $\lambda _{m,n}\geq0$ and ${{\alpha _n}}\geq0$. Then, the Lagrangian of $\mathbf{P}_{{2}}$ can be given by $\left(27\right)$ at the tope of this page,
\begin{figure*}[!t]
\normalsize
\begin{align}\label{27}\ \notag
{\cal {L}}\left( \Xi  \right){\rm{ = }}&\sum\limits_{m = 1}^M {{w_m}\left[ {\sum\limits_{n = 1}^N {\frac{{{f_m}\left[ n \right]}}{C}} \frac{T}{N} + \frac{{BT{t_m}\left[ n \right]}}{{{\nu _m}N}}{{\log }_2}\left( {1 + \frac{{{h_m}\left[ n \right]{z_m}\left[ n \right]}}{{\sigma _0^2{t_m}\left[ n \right]}}} \right)} \right]} \\
&{\rm{ + }}\sum\limits_{m = 1}^M {\sum\limits_{n = 1}^N {{\lambda _{m,n}}} } \left\{ {\frac{{{\eta _0}T}}{N}\sum\limits_{k = 1}^n {{h_m}\left[ k \right]{P_0} - \frac{T}{N}\sum\limits_{k = 1}^n {\left[ {{\gamma _c}f_m^3\left[ k \right] + {z_m}\left[ k \right]} \right]} } } \right\}+ \sum\limits_{n = 1}^N {{\alpha _n}\left\{ {1 - \sum\limits_{m = 1}^M {{t_m}\left[ n \right]} } \right\}},
\end{align}
\hrulefill \vspace*{4pt}
\end{figure*}
where $\Xi $ denotes a collection of all the  primal and dual variables related to $\mathbf{P}_{{2}}$. Let ${\mu _{m,n}} = \sum\limits_{k = n}^N {{\lambda _{m,k}}}$ and ${g_m}\left[ k \right] = {\eta _0}{h_m}\left[ k \right]P_0 - {\gamma _c}f_m^3\left[ k \right] - {z_m}\left[ k \right]$. Then, the Lagrangian function $L\left( \Xi  \right){\rm{  }}$ can be rewritten  by $\left(28\right)$ at the tope of this page.
\begin{figure*}[!t]
\normalsize
\begin{align}\label{27}\
{\cal {L}}\left( \Xi  \right) = \sum\limits_{m = 1}^M {\sum\limits_{n = 1}^N {{w_m}\left\{ {\frac{T}{N}\frac{{{f_m}\left[ n \right]}}{C} + \frac{{BT{t_m}\left[ n \right]}}{{{\nu _m}N}}{{\log }_2}\left( {1 + \frac{{{h_m}\left[ n \right]{z_m}\left[ n \right]}}{{\sigma _0^2{t_m}\left[ n \right]}}} \right)} \right\}{\rm{ + }}{\mu _{m,n}}{g_m}\left[ k \right]{\rm{ + }}\frac{{{\alpha _n}}}{M} - {\alpha _n}{t_m}\left[ n \right]} }.
\end{align}
\hrulefill \vspace*{4pt}
\end{figure*}
And the Lagrangian dual function of $\mathbf{P}_{{2}}$ can be presented as
\begin{align}\label{27}\
g\left( {\lambda _{m,n} ,\alpha _n } \right) = \begin{array}{*{20}{c}}
{\mathop {\max }\limits_{0 \le {f_m}\left[ n  \right] } }&{{\cal {L}}\left( \Xi \right)}.
\end{array}
\end{align}
Based on $\left(29\right)$, the optimal solutions of $\mathbf{P}_{{2}}$ can be obtained by solving its dual problem, given as
\begin{align}\label{27}\
\mathop {\min }\limits_{{\lambda _{m,n}},{\alpha _n}} g\left( {{\lambda _{m,n}},{\alpha _n}} \right).
\end{align}
It can be seen from $\left(30\right)$ that the dual problem can be  decoupled into $M$ independent optimization problems, given by $\left(31\right)$ at the tope of the next page.
\begin{figure*}[!t]
\normalsize
\begin{subequations}
\begin{align}\label{27}\
&\mathop {\max }\limits_{{\lambda _{m,n}},{\alpha _n},{f_m}\left[ n \right] \ge 0} {{\cal {L}}_{m}}\left( {{\lambda _{m,n}},{\alpha _n},{f_m}\left[ n \right],{z_m}\left[ n \right],{t_m}\left[ n \right]} \right)\\ \notag \\
&{{\cal {L}}_{m}}\left( {{\lambda _{m,n}},{\alpha _n},{f_m}\left[ n \right],{z_m}\left[ n \right],{t_m}\left[ n \right]} \right)\\
 &\ \ \ \ \ = \sum\limits_{n = 1}^N {\left\{ {{w_m}\left\{ {\frac{T}{N}\frac{{{f_m}\left[ n \right]}}{C} + \frac{{BT{t_m}\left[ n \right]}}{{{\nu _m}N}}{{\log }_2}\left( {1 + \frac{{{h_m}\left[ n \right]{z_m}\left[ n \right]}}{{\sigma _0^2{t_m}\left[ n \right]}}} \right)} \right\}{\rm{ + }}{\mu _{m,n}}{g_m}\left[ n \right]{\rm{ + }}\frac{{{\alpha _n}}}{M} - {\alpha _n}{t_m}\left[ n \right]} \right\}} .
\end{align}
\end{subequations}
\hrulefill \vspace*{4pt}
\end{figure*}
Thus, let the derivation of $\left(31\rm{b}\right)$ with respect to ${f_m}\left[ n \right]$ and${z_m}\left[ n \right]$ be zero, one has
\begin{subequations}
\begin{align}\label{27}\
&\frac{{T{w_m}}}{{NC}} - \frac{{3T{\gamma _c}f_m^2\left[ k \right]}}{N}\sum\limits_{k = n}^N {{\lambda _{m,k}}}  = 0,\\
&\frac{{{w_m}BT{t_m}\left[ n \right]}}{{{\nu _m}N\ln 2}}\frac{{{h_m}\left[ n \right]}}{{\sigma _0^2{t_m}\left[ n \right] + {h_m}\left[ n \right]{z_m}\left[ n \right]}} - \frac{T}{N}\sum\limits_{k = n}^N {{\lambda _{m,k}}} {\rm{ = }}0.
\end{align}
\end{subequations}
Note that ${z_m}\left[ k \right] = {t_m}\left[ k \right]{P_m}\left[ k \right]$ and ${P_m}\left[ k \right]\geq0$. Moreover, the case that ${t_m}\left[ n \right]=0$ can be identified as ${P_m}\left[ n \right]=0$. Thus, based on $\left(32\right)$, Theorem 1 is proved. The proof for Theorem 1 is complete.

\section{Proof of Theorem 5}
Let ${{\upsilon _{m,n}}}$ and ${{\varepsilon _n}}$ denote the dual variables with respect to the constraints given by $\left(24\rm{b}\right)$ and $\left(24\rm{c}\right)$, respectively, where ${{\upsilon _{m,n}}}\geq0$ and ${{\varepsilon _n}}\geq0$. Then, for any given $f_m\left[ n \right]$, ${P_m}\left[ n \right]$, $t_m\left[ n \right]$ and $\mathbf{q}_u\left[ n \right]$, the Lagrangian of $\mathbf{P}_{{6}}$ can be expressed by $\left(33\right)$ at the tope of the next page,
\begin{figure*}[!t]
\normalsize
\begin{align}\label{27}\ \notag
{\cal {L}}_1\left( \Xi_1  \right){\rm{ = }}&\sum\limits_{m = 1}^M {{w_m}\left[ {\left( {1 - {\rho _m}} \right)\sum\limits_{n = 1}^N {\frac{{{f_m}\left[ n \right]}}{C}} \frac{T}{N} + \frac{{BT{\rho _m}{t_m}\left[ n \right]}}{{{\nu _m}N}}{{\log }_2}\left( {1 + \frac{{{h_m}\left[ n \right]{z_m}\left[ n \right]}}{{{t_m}\left[ n \right]\sigma _0^2}}} \right)} \right]} \\ \notag
&{\rm{ + }}\sum\limits_{m = 1}^M {\sum\limits_{n = 1}^N {{\upsilon _{m,n}}} } \left\{ {\frac{{{\eta _0}T}}{N}\sum\limits_{k = 1}^n {{h_m}\left[ k \right]{P_0} - \frac{T}{N}\sum\limits_{k = 1}^n {\left[ {\left( {1 - {\rho _m}} \right){\gamma _c}f_m^3\left[ k \right] + {\rho _m}{z_m}\left[ k \right]} \right]} } } \right\}\\
 &+ \sum\limits_{n = 1}^N {{\varepsilon _n}\left\{ {1 - \sum\limits_{m = 1}^M {{\rho _m}{t_m}\left[ n \right]} } \right\}},
\end{align}
\hrulefill \vspace*{4pt}
\end{figure*}
where $\Xi_1 $ denotes a collection of all the  primal and dual variables related to $\text{P}_{{6}}$. $\Xi_2 $ denotes a collection of ${\upsilon _{m,n}}$, ${\alpha _n}$, ${f_m}\left[ n \right]$ , ${z_m}\left[ n \right]$, ${t_m}\left[ n \right]$ and ${\rho _m}$. Using the same techniques that are used for the proof of Theorem 1, for any given $f_m\left[ n \right]$, ${z_m}\left[ n \right]$, $t_m\left[ n \right]$ and $\mathbf{q}_u\left[ n \right]$, $\mathbf{P}_{{6}}$ can be solved by solving $M$  independent optimization problems, given by $\left(34\right)$ at the tope of the next page,
\begin{figure*}[!t]
\normalsize
\begin{subequations}
\begin{align}\label{27}\
&\mathop {\max }\limits_{{{{\upsilon _{m,n}}}},{\varepsilon _n},{f_m}\left[ n \right] \ge 0} {{\cal {L}}_{m}^1}\left( \Xi_2 \right)\\ \notag
&{{\cal {L}}_{m}^1}\left( \Xi_2  \right)= \sum\limits_{n = 1}^N { {{w_m}\left\{ {\frac{{T\left( {1 - {\rho _m}} \right){f_m}\left[ n \right]}}{NC} + \frac{{BT{\rho _m}{t_m}\left[ n \right]}}{{{\nu _m}N}}{{\log }_2}\left( {1 + \frac{{{h_m}\left[ n \right]{z_m}\left[ n \right]}}{{{t_m}\left[ n \right]\sigma _0^2}}} \right)} \right\}{\rm{  }}} }\\
 &\ \ \ \ \ \ \ \ +\sum\limits_{n = 1}^N { {\varpi _{m,n}}{\ell_m}\left[ n \right]{\rm{ + }}\frac{{{\varepsilon _n}}}{M} - {\varepsilon _n}{t_m}\left[ n \right]},
\end{align}
\end{subequations}
\hrulefill \vspace*{4pt}
\end{figure*}
where ${\ell_m}\left[ n \right]= {\eta _0}{h_m}\left[ n \right]P_0 - \left( {1 - {\rho _m}} \right){\gamma _c}f_m^3\left[ n \right] - {\rho _m}{z_m}\left[ n \right]$ and $\varpi _{m,n}=\sum\limits_{k = n}^N {{\upsilon _{m,k}}}$. Thus, according to \cite{F. Zhou}, the optimal ${\rho _m}$ denoted by ${\rho _m^{opt}}$ can be obtained by $\left(35\right)$ at the tope of the next page.
\begin{figure*}[!t]
\normalsize
\begin{subequations}
\begin{align}\label{27}\
&\frac{{\partial {{\cal {L}}_{m}^1}\left( \Xi_2  \right)}}{{\partial \rho _{m}^{opt}}}\left\{ \begin{array}{l}
\begin{array}{*{20}{c}}
{ < 0,}&{\rho _{m}^{opt} = 0,}
\end{array}\\
\begin{array}{*{20}{c}}
{ = 0,}&{0 < \rho _{m}^{opt} < 1,}
\end{array}\\
\begin{array}{*{20}{c}}
{ > 0,}&{\rho _{m}^{opt} = 1;}
\end{array}
\end{array} \right. m\in {\cal M} \\ \notag
& \frac{{\partial {{\cal {L}}_{m}^1}\left( \Xi_2  \right)}}{{\partial \rho _{m}^{opt}}} = \left\{ {\sum\limits_{n = 1}^N { - \frac{{{w_m}{f_m}\left[ n \right]}}{C}} \frac{T}{N} + \frac{{BT{t_m}\left[ n \right]}}{{{\nu _m}N}}{{\log }_2}\left( {1 + \frac{{{h_m}\left[ n \right]{z_m}\left[ n \right]}}{{{t_m}\left[ n \right]\sigma _0^2}}} \right)} \right\} \\
&\ \ \ \ \ \ \ \ \ \ \ \ \ \ +  \sum\limits_{n = 1}^N {{\upsilon _{m,n}}} \left\{ { - \frac{T}{N}\sum\limits_{k = 1}^n {\left[ { - {\gamma _c}f_m^3\left[ k \right] + {z_m}\left[ k \right]} \right]} } \right\} - \sum\limits_{n = 1}^N {{\varepsilon _n}{t_m}\left[ n \right]}.
\end{align}
\end{subequations}
\hrulefill \vspace*{4pt}
\end{figure*}
Based on $\left(35\right)$, since ${z_m}\left[ n \right]={t_m}\left[ n \right]{P_m}\left[ n \right]$, Theorem 5 is proved.

\begin{IEEEbiography}[{\includegraphics[width=1.0in,height=1.15in,clip,keepaspectratio]{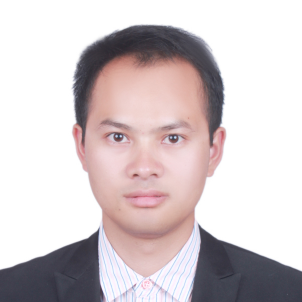}}]{Fuhui Zhou}

received the Ph. D. degree from Xidian University, Xi¡¯an, China, in 2016. He is an associate Professor with School of Information Engineering, Nanchang University. He is now a Research Fellow at Utah State University. He has worked as an international visiting Ph. D student of the University of British Columbia from 2015 to 2016. His research interests focus on cognitive radio, green communications, edge computing, machine learning, NOMA, physical layer security, and resource allocation. He has published more than 40 papers, including IEEE Journal of Selected Areas in Communications, IEEE Transactions on Wireless Communications, IEEE Wireless Communications, IEEE Network, IEEE GLOBECOM, etc. He has served as Technical Program Committee (TPC) member for many International conferences, such as IEEE GLOBECOM, IEEE ICC, etc. He serves as an Associate Editor of IEEE Access.
\end{IEEEbiography}

\begin{IEEEbiography}[{\includegraphics[width=1.05in,height=1.3in,clip,keepaspectratio]{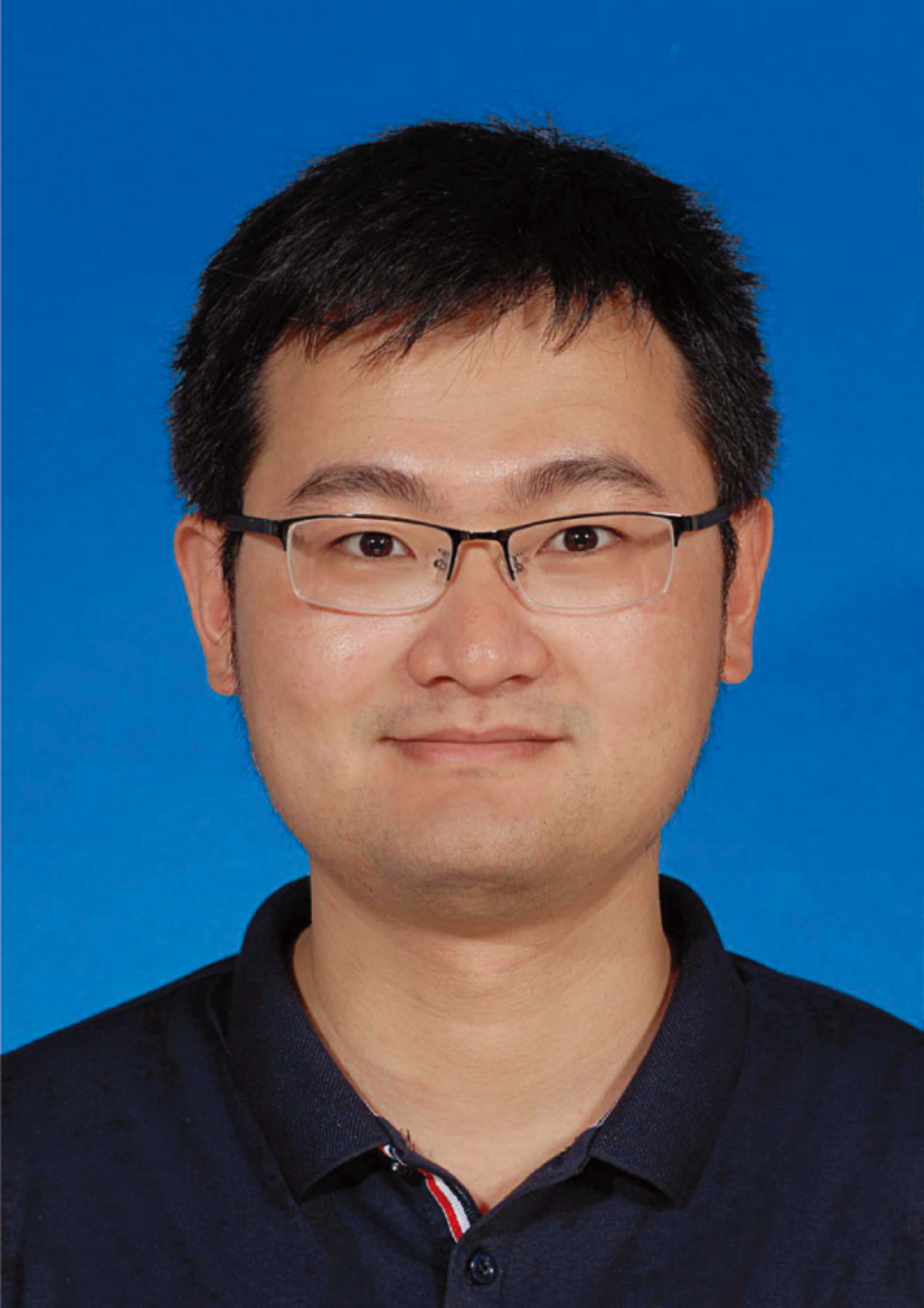}}]{Yongpeng Wu}(S'08--M'13--SM'17)
received the B.S. degree in telecommunication engineering from Wuhan University, Wuhan, China, in July 2007, the Ph.D. degree in communication and signal processing with the National Mobile Communications Research Laboratory, Southeast University, Nanjing, China, in November 2013.

Dr. Wu is currently a Tenure-Track Associate Professor with the Department of Electronic Engineering, Shanghai Jiao Tong University, China. Previously, he was senior research fellow with Institute for Communications Engineering, Technical University of Munich, Germany and the Humboldt research fellow and the senior research fellow with Institute for Digital Communications, University Erlangen-N$\ddot{u}$rnberg, Germany. During his doctoral studies, he conducted cooperative research at the Department of Electrical Engineering, Missouri University of Science and Technology, USA.
His research interests include massive MIMO/MIMO systems, physical layer security, signal processing for wireless communications, and multivariate statistical theory.

Dr. Wu was awarded the IEEE Student Travel Grants for IEEE International Conference on Communications (ICC) 2010, the Alexander von Humboldt Fellowship in 2014, the Travel Grants for IEEE Communication Theory Workshop 2016, and the Excellent Doctoral Thesis Awards of China Communications Society 2016. He was an Exemplary Reviewer of the IEEE Transactions on Communications in 2015, 2016.
He is the lead guest editor for the upcoming special issue ``Physical Layer Security for 5G Wireless Networks" of the IEEE Journal on Selected Areas in Communications.
He is currently an editor of the IEEE Access and IEEE Communications Letters. He has been a TPC member of various conferences, including Globecom, ICC, VTC, and PIMRC, etc.
\end{IEEEbiography}

\begin{IEEEbiography}[{\includegraphics[width=1.0in,height=2.15in,clip,keepaspectratio]{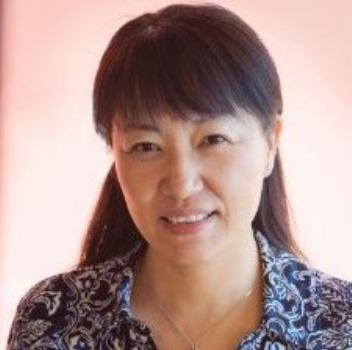}}]{Rose Qingyang Hu}

is a Professor of Electrical and Computer Engineering Department at Utah State University. She received her B.S. degree from University of Science and Technology of China, her M.S. degree from New York University, and her Ph.D. degree from the University of Kansas. She has more than 10 years of R\&D experience with Nortel, Blackberry and Intel as a technical manager, a senior wireless system architect, and a senior research scientist, actively participating in industrial 3G/4G technology development, standardization, system level simulation and performance evaluation. Her current research interests include next-generation wireless communications, wireless system design and optimization, green radios, Internet of Things, Cloud computing/fog computing, multimedia QoS/QoE, wireless system modeling and performance analysis. She has published over 180 papers in top IEEE journals and conferences and holds numerous patents in her research areas. Prof. Hu is an IEEE Communications Society Distinguished Lecturer Class 2015-2018 and the recipient of Best Paper Awards from IEEE Globecom 2012, IEEE ICC 2015, IEEE VTC Spring 2016, and IEEE ICC 2016.
\end{IEEEbiography}

\begin{IEEEbiography}[{\includegraphics[width=1.0in,height=2.15in,clip,keepaspectratio]{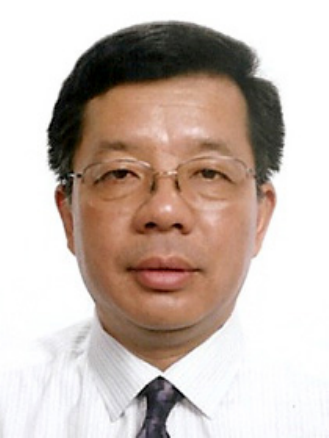}}]{Yi Qian}

received a Ph.D. degree in electrical engineering from Clemson University. He is a professor in the Department of Electrical and Computer Engineering, University of Nebraska-Lincoln (UNL). Prior to joining UNL, he worked in the telecommunications industry, academia, and the government. Some of his previous professional positions include serving as a senior member of scientific staff and a technical advisor at Nortel Networks, a senior systems engineer and a technical advisor at several start-up companies, an assistant professor at University of Puerto Rico at Mayaguez, and a senior researcher at National Institute of Standards and Technology. His research interests include information assurance and network security, network design, network modeling, simulation and performance analysis for next generation wireless networks, wireless ad-hoc and sensor networks, vehicular networks, smart grid communication networks, broadband satellite networks, optical networks, high-speed networks and the Internet. Prof. Yi Qian is a member of ACM and a senior member of IEEE. He was the Chair of IEEE Communications Society Technical Committee for Communications and Information Security from January 1, 2014 to December 31, 2015. He is a Distinguished Lecturer for IEEE Vehicular Technology Society and IEEE Communications Society. He is serving on the editorial boards for several international journals and magazines, including serving as the Associate Editor-in-Chief for IEEE Wireless Communications Magazine. He is the Technical Program Chair for IEEE International Conference on Communications (ICC) 2018.
\end{IEEEbiography}


\begin{thebibliography}{20}
\bibitem{F. Zhou4}
F. Zhou, Y. Wu, R. Q. Hu, Y. Wang, and K. K. Wong, \lq\lq Energy-efficient NOMA enabled  heterogeneous cloud radio access networks,\rq\rq \ \emph{IEEE Network}, vol. 32, no. 2, pp.152-160, 2018.
\bibitem{Y. Mao}
Y. Mao, C. You, J. Zhang, K. Huang, and K. B. Letaief, \lq\lq A survey on mobile edge computing: The communication perspective,\rq\rq \ \emph{IEEE Commun. Surveys Tuts.}, vol. 19, no. 4, pp. 2322-2358, Fourth Quarter, 2017.
\bibitem{X. Lu}
X. Lu, P. Wang, D. Niyato, D. I. Kim, and Z. Han, \lq\lq Wireless networks with RF energy harvesting: A contemporary survey,\rq\rq \ \emph{IEEE Commun. Surveys Tuts.}, vol. 17, pp. 757-789, Second Quarter, 2015.
\bibitem{F. Zhou2}
F. Zhou, Z. Li, J. Cheng, Q. Li, and J. Si, \lq\lq Robust AN-aided beamforming and power splitting design for secure MISO cognitive radio with SWIPT,\rq\rq \ \emph{IEEE Trans. Wireless Commun.}, vol. 16, no. 4, pp. 2450-2464, April 2017.
\bibitem{S. Sardellitti}
S. Sardellitti, G. Scutari, and S. Barbarossa, \lq\lq Joint optimization of radio and computational resources for multicell mobile-edge computing,\rq\rq \ \emph{IEEE Trans. Signal Inf. Process. Over Netw.}, vol. 1, no. 2, pp. 89-103, Jun. 2015.
\bibitem{C. You1}
C. You, K. Huang, H. Chae, and B. Kim, \lq\lq Energy-efficient resource allocation for mobile-edge computation offloading,\rq\rq \ \emph{IEEE Trans. Wireless Commun.}, vol. 16, no. 3, pp. 1397-1411, Mar. 2017.
\bibitem{C. Wang}
C. Wang, C. Liang, F. R. Yu, Q. Chen, L. Tang, \lq\lq Computation offloading and resource allocation in wireless cellular networks with moible edge computing,\rq\rq \ \emph{IEEE Trans. Wireless Commun.}, vol. 16, no. 8, pp. 4924-4938, Aug. 2017.
\bibitem{J. Du}
J. Du, L. Zhao, J. Feng, and X. Chu, \lq\lq Computation offloading and resource allocation in mixed fog/cloud computing systems with min-max fairness guarantee,\rq\rq \ \emph{IEEE Trans. Commun.}, vol. 66, no. 4, pp. 1594-1608, April 2018.
\bibitem{L. Liu}
L. Liu, Z. Chang, X. Guo, S. Mao, and T. Ristaniemi, \lq\lq Multi-objective optimization for computation offloading in fog computing,\rq\rq \ \emph{IEEE Internet Things J.}, vol. 5, no. 1, pp. 283-294, Jan. 2018.
\bibitem{W. Zhang}
W. Zhang, Y. Wen, K. Guan, D. Kilper, H. Luo, and D. O. Wu, \lq\lq Energy-optimal mobile cloud computing under stochastic wireless channel,\rq\rq \ \emph{IEEE Trans. Wireless Commun.}, vol. 12, no. 9, pp. 4569-4581, Sep. 2013.
\bibitem{J. Xu}
J. Xu, L. Chen, and S. Ren, \lq\lq Online learning for offloading and autoscaling in energy harvesting mobile edge computing,\rq\rq \ \emph{IEEE Trans. Cogn. Netw.}, vol. 3, no. 3, pp. 361-373, Sep., 2017.
\bibitem{Y. Mao1}
Y. Mao, J. Zhang, and K. B. Letaief, \lq\lq Dynamic computation offloading for mobile-edge computing with energy harvesting devices,\rq\rq \ \emph{IEEE J. Sel. Areas Commun.}, vol. 34, no. 12, pp. 3590-3605, Dec., 2016.
\bibitem{C. You}
C. You, K. Huang, and H. Chae, \lq\lq Energy efficient mobile cloud computing powered by wireless energy transfer,\rq\rq \ \emph{IEEE J. Sel. Areas Commun.}, vol. 34, no. 5, pp. 1757-1771, May, 2016.
\bibitem{F. Wang}
F. Wang, J. Xu, X. Wang, and S. Cui, \lq\lq Joint offloading and computing optimization in wireless powered mobile-edge computing systems,\rq\rq \ \emph{IEEE Trans. Wireless Commun.}, vol. 17, no. 3, pp. 1784-1797, March 2018.
\bibitem{S. Mao}
S. Mao, S. Leng, K. Yang, X. Huang, and Q. Zhao, \lq\lq Fair energy-efficient scheduling in wireless powered full-duplex mobile-egde computing systems,\rq\rq \ in \emph{Proc. IEEE Global Commun. Conf.}, Singapore, 2017.
\bibitem{S. Bi}
S. Bi and Y. Zhang, \lq\lq Computation rate maximization for wireless powered mobile egde computing with binary computation offloading,\rq\rq \  \emph{IEEE Trans. Wireless Commun.}, to be published, 2018.
\bibitem{H. Wang}
H. Wang, J. Wang, G. Ding, L. Wang, T. A. Tsiftsis, P. K. Sharma, \lq\lq Resource allocation for energy harvesting-powered D2D communication underlaying UAV-assisted networks,\rq\rq \ \emph{IEEE Trans. Cogn. Netw.}, vol. 2, no. 1, pp. 14-24, Jan. 2018.
\bibitem{S. Yin}
S. Yin, J. Tan, and L. Li, \lq\lq UAV-assisted cooperative communications with wireless information and power transfer,\rq\rq \ submitted to \emph{IEEE Trans. Wireless Commun.}, https://arxiv.org/abs/1710.00174v1.
\bibitem{J. Xu2}
J. Xu, Y. Zeng, and R. Zhang, \lq\lq UAV-enabled wireless power transfer: Trajectory design and energy region charaterization,\rq\rq \ in \emph{Proc. IEEE Global Commun. Conf.} Singapore, 2017,
\bibitem{J. Xu3}
J. Xu, Y. Zeng, and R. Zhang, \lq\lq UAV-enabled wireless power transfer: Trajectory design and energy optimization,\rq\rq \  \emph{IEEE Trans. Wireless Commun.}, to be published, 2018.
\bibitem{N. H. Motlagh}
N. H. Motlagh, M. Bagaa, and T. Taleb, \lq\lq UAV-based IoT platform: A crowd surveillance use case,\rq\rq \ \emph{IEEE Commun. Mag.}, vol. 55, no. 2, pp. 128-134, Feb. 2017.
\bibitem{N. Zhao}
N. Zhao, F. Cheng, F. R. Yu, J. Tang, Y. Chen, G. Gui, and H. Sari, \lq\lq Caching UAV assisted secure transmission in hyper-dense networks based on interference alignment,\rq\rq \ \emph{IEEE Trans. Commun.}, vol. 66, no. 5, pp. 2281-2294, May 2018.
\bibitem{S. Jeong}
S. Jeong, O. Simeone, and J. Kang, \lq\lq Mobile edge computing via a UAV-mounted cloudlet: Optimization of bit allocation and path planning,\rq\rq \ \emph{IEEE Trans. Vehicular Technol.}, vol. 67, no. 3, pp. 2049-2063, Mar. 2018.
 \bibitem{S. Jeong2}
S. Jeong, O. Simeone, and J. Kang, \lq\lq Mobile edge computing with a UAV-mounted cloudlet: Optimal bit allocation for communication and computation,\rq\rq \ \emph{IET Commmn.}, vol. 11, no. 7, pp. 969-974, Nov. 2017.
\bibitem{M. A. Messous}
M. A. Messous, H. Sedjelmaci, N. Houari, and S. M. Senouci, \lq\lq Computation offloading game for an UAV network in mobile egde computing,\rq\rq \ in\ \emph{Proc. IEEE Int. Conf. Commuun.}, France, May, 2017.
\bibitem{Y. Zeng1}
Y. Zeng, R. Zhang, and T. J. Lim, \lq\lq Wireless communications with unmanned aerial vehicles: Opportunities and challenges,\rq\rq \ \emph{IEEE Commun. Mag.}, vol. 54, no. 5, pp. 36-42, May 2016.
\bibitem{Y. Zeng}
Y. Zeng and R. Zhang, \lq\lq Energy-efficient UAV communication with trajectory optimization,\rq\rq \ \emph{IEEE Trans. Wireless Commun.}, vol. 16, no. 6, pp. 3747-3760, June 2017.
\bibitem{P. Yang}
P. Yang, X. Cao, C. Yin, Z. Xiao, X. Xi, and D. Wu, \lq\lq Proactive drone-cell deployment: Overload relief for a cellular network under flash crowd traffic,\rq\rq \ \emph{IEEE Trans. Intell. Transportation Sys.}, vol. 18, no. 10, pp. 2877-2892, Oct, 2017.
\bibitem{E. Kalantari}
E. Kalantari, H. Yanikomeroglu, and A. Yongacoglu, \lq\lq On the number and 3D placement of drone base stations in wireless cellular networks,\rq\rq in \ \emph{Proc. IEEE VTC fall, 2016}, Montreal, Canada, Sept. 2016.
\bibitem{E. Kalantari1}
E. Kalantari, M. Z. Shakir, H. Yanikomeroglu, and A. Yongacoglu, \lq\lq Backhaul-aware robust 3D drone placement in 5G+ wireless networks,\rq\rq in \ \emph{Proc. ICC Workshops, 2017}, Paris, France, May 2017.
\bibitem{L. Zeng}
L. Zeng, X. Cheng, C. X. Wang, and X. Yin, \lq\lq A 3D geometry-based stochastic channel model for UAV-MIMO channels,\rq\rq \ in \ \emph{Proc. IEEE WCNC 2017}, San Francisco, USA, Mar. 2017.
\bibitem{C. X. Wang}
C. X. Wang, A. Ghazal, B. Ai, P. Fan, and Y. Liu, \lq\lq Channel measurements and models for high-speed train communication systems: a survey,\rq\rq \ \emph{IEEE Commun. Surveys Tuts.}, vol. 18, no. 2, pp. 974-987, 2nd Quart., 2016.
\bibitem{N. Cheng}
N. Cheng, W. Xu, W. Shi, Y. Zhou, N. Lu, H. Zhou, and X. Shen, \lq\lq Aire-ground integrated mobile edge networks: Architecture, challenges and opportunities,\rq\rq \ \emph{IEEE Commun. Mag.}, to be published, 2018.
\bibitem{M. Mozaffari}
M. Mozaffari, W. Saad, M. Bennis, Y. H. Nam, and M. Debbah, \lq\lq A tutorial on UAVs for wireless networks: Applications, challenges, and open problems,\rq\rq \ \emph{IEEE Commun. Surveys Tuts.}, submitted, 2018. https://arxiv.org/abs/1803.00680
\bibitem{S. P. Boyd}
S. P. Boyd and L. Vandenberghe, \emph{Convex Optimization}. Cambridge, U.K.: Cambridge Univ. Press, 2004.
\bibitem{F. H. Zhou}
F. Zhou, N. C. Beaulieu, Z. Li, J. Si, and P. Qi, \lq\lq Energy-efficient optimal power allocation for fading cognitive radio channels: Ergodic capacity, outage capacity and minimum-rate capacity,\rq\rq \ \emph{IEEE Trans. Wireless Commun.}, vol. 15, no. 4, pp. 2741-2755, Apr. 2016.
\bibitem{F. Zhou}
F. Zhou, Z. Li, J. Cheng, Q. Li, and J. Si, \lq\lq Robust max-min fairness resource allocation in sensing-based wideband cognitive radio with SWIPT: Imperfect channel sensing,\rq\rq \ \emph{IEEE Syst. J.}, to be published, 2017.
\bibitem{S. Bubeck}
S. Bubeck,  \lq\lq Convex optimization: Algorithms and complexity,\rq\rq \ \emph{In Foundations and Trends in Machine Learning}, vol. 8, no. 3, pp. 231-357, 2015. https://arxiv.org/abs/1405.4980
\bibitem{C. Gutierrez}
C. Gutierrez, F. Gutierrez, M.C. Rivara, \lq\lq Complexity on the bisection method,\rq\rq \ \emph{Theoretical Computer Science}, vol. 382, pp. 131-138, 2007.
\end{thebibliography}
\end{document}